# Determination of the roles of strain and tearing in single photon emission from nanoindented $WSe_2$

Joseph Stage[1,†], John Pierce Fix[1,2,†], Matthew C. Strasbourg[3], Amir Darabi[1], Torrey McLoughlin[1], Sheikh Parvez[1,2], Elijah Stuvland[1], Andrew R. Lingley[4], Thomas Darlington[5], and Nicholas J. Borys[1,6,*]

[1]Department of Physics, Montana State University, Bozeman, MT, 59717, USA

[2]Materials Science and Engineering, Montana State University, Bozeman, MT, 59717, USA

[3]Department of Mechanical Engineering, Columbia University, New York City, NY, 10027, USA

[4]Department of Electrical and Computer Engineering, Montana State University, Bozeman, MT, 59717, USA

[5]Molecular Foundry, Lawrence Berkeley National Laboratory, Berkeley, CA, 94720, USA

[6]Department of Physics and Astronomy, University of Utah, Salt Lake City, UT, 84109, USA

**Abstract:** Single-photon emission in two-dimensional single-layer $WSe_2$ is attractive for the on-demand generation of quantum states of light. The electronic states that are responsible for single-photon generation preferentially form in regions of localized tensile strain, enabling deterministic positioning and strain engineering. Nanoindentation of single-layer $WSe_2$ yields controlled and reproducible deformations that generate the localized strain needed to activate the single-photon-emitting states. However, using nanoindentation both for investigating structure-property relationships and for manufacturing quantum light sources based on $WSe_2$ is hindered by key questions on the structural integrity of the indented 2D material, the resulting strain generated, and the sub-micron location of the emitters. In this work, we study the structure of indented single-layer $WSe_2$ using a fabrication process that inverts the indents into protruding pillars that can be probed using electron microscopy. We explicitly identify strain relaxation of the indented single-layer $WSe_2$ due to tearing and confirm that single-photon-emitting states still form in these systems, likely at the extremities of the tear. For indents that are confirmed to be intact (i.e., not torn), we assess the ability to strain engineer the single-photon emitters. While strain does not strongly affect the emission energy or the brightness, we find that increased strain reduces the spatial density of emitters. This trend indicates that an optimal amount of strain is needed for emitter formation and/or the emitters preferentially form on the periphery of the indent. Our investigation provides insight into the most relevant structure-property relationships for using strain to engineer quantum light sources in single-layer $WSe_2$ and other 2D semiconductors.

---

[†] J. S. and J. P. F. contributed equally to this work
[*] Corresponding author: nicholas.borys@utah.edu

## Introduction

From new developments to improvements in performance, many aspects of quantum technologies stand to benefit from a system that produces tailorable quantum states of light on demand. Solid-state single-photon emitters (SPEs) that are based on localized electronic states in semiconducting and insulating materials have the potential to address this need, and further, have several advantageous physical properties for technological development[1-5]. One appealing opportunity for this class of SPEs is their ultimate development into an electrically triggerable device with a micron or sub-micron footprint that is compatible with photonic integrated circuits. Solid-state SPE systems that are ideal for such device integration, however, need to meet stringent requirements. Not only do they need to be able to produce single photons at sufficient purity and indistinguishability, but they also need to be amenable to advanced manufacturing and nanofabrication. In particular, they need to possess the versatility to be deterministically positioned and heterogeneously integrated with photonic and nanophotonic systems such as waveguides and cavities.

Several aspects of the localized states responsible for single-photon emission in single-layer $WSe_2$ (1L-$WSe_2$)—a prototypical 2D semiconductor—make them appealing for solid-state SPE devices. Since their initial discovery in 1L-$WSe_2$[6-13], prototype electrically triggerable SPE devices have been demonstrated[13-17], strain-engineering has been explored[10, 18-21], and non-classical emission has been reported at temperatures up to 160 K[22, 23]. Additionally, it is well established that the SPE-active state preferentially forms in regions of localized tensile strain[4, 10, 19, 24, 25], a property that can be leveraged to achieve deterministic placement of the SPE state[14, 26-34]. Nanoindentation, where a nanoscale atomic force microscopy (AFM) probe is used to plastically indent 1L-$WSe_2$ on top of a deformable polymer, is an elegant method to deterministically position regions of localized tensile strain as well as tailor the amount of strain with the indentation force. Following the first report of creating SPE states with nanoindentation[32], the technique has been leveraged to demonstrate how graphite can improve photon purity[28], as well as deterministically place an SPE state in an electrically triggered device[14] and near a plasmonic structure[35]. Despite these initial successes, several questions about the nanoindentation procedure and the extent to which SPEs can be strain-engineered persist. Specifically, the roles, whether deleterious or advantageous, of micro- and nanoscale tearing are unknown, and the properties of the SPEs appear to only weakly depend on the amount of strain in the indent. Further, it is unclear where within the indented structure the SPE states exist[26]. These unresolved issues are critical impediments to achieving reproducible manufacturing of SPE states in 2D semiconductors using nanoindentation and also apply to understanding SPE creation/engineering in other approaches to deterministically generate localized strain.

Here, we report a detailed study of the structure of nanoindented 1L-$WSe_2$ that resolves the presence of tearing. Using photoluminescence (PL) spectroscopy and imaging, we confirm that tearing can occur and relax a large amount of the strain in the indented structure. Despite the strain relaxation, we confirm that SPEs form at the indents where tearing occurs, which strongly suggests that, in these cases, the SPE states are localized to the tear. Further, we use our ability to discern tearing to assess the ability to strain-engineer the SPE states in indents that remain intact and preserve the localized strain. By explicitly eliminating the instances where tearing occurs, we find that increasing the amount of strain with the indentation force does not significantly alter the physical properties of the generated SPE states themselves. However, we do find that as the indentation force increases, the spatial density of the SPE states decreases, which suggests that either an optimal range of tensile strain exists for the formation of SPE states or that the SPE states preferentially localize to the periphery of the indent, as was concluded from super-resolution imaging measurements[26]. Our results substantially improve the understanding of the SPE formation process in indented 1L-$WSe_2$ and, as a result, also improve the ability to leverage nanoindentation as a highly controllable process to position and strain-engineer SPEs in all 2D systems.

**Results and Discussion**

Figure 1a illustrates the process of generating the localized strain in the 1L-$WSe_2$ using nanoindentation[14, 26-34, 36]. The samples are comprised of 1L-$WSe_2$ that is directly exfoliated onto ~350 nm thick PMMA films. To indent the 1L-$WSe_2$, an AFM probe is brought into initial contact at a fixed location on the 1L-$WSe_2$/PMMA layer. The contact force can then be increased by further raising the sample into the AFM probe, where, at sufficiently large forces, the PMMA is plastically deformed into an indent. Because of strong adhesion, the 1L-$WSe_2$ is suspected to be conformally stretched along the profile of the indent, resulting in a region of localized strain where the amount of strain is dependent upon the size (e.g., depth) of the indent. The geometry/structure of the indent, and thus the generated strain, depends on the nano- and microscale structure of the AFM probe combined with the displacement of the PMMA, resulting in a peripheral buildup region around the indent. Further, the depth, width, and size of the buildup region for a given indent depend on the force constant of the AFM cantilever used during indentation and the additional $z$-displacement of the sample beyond the initial contact point.

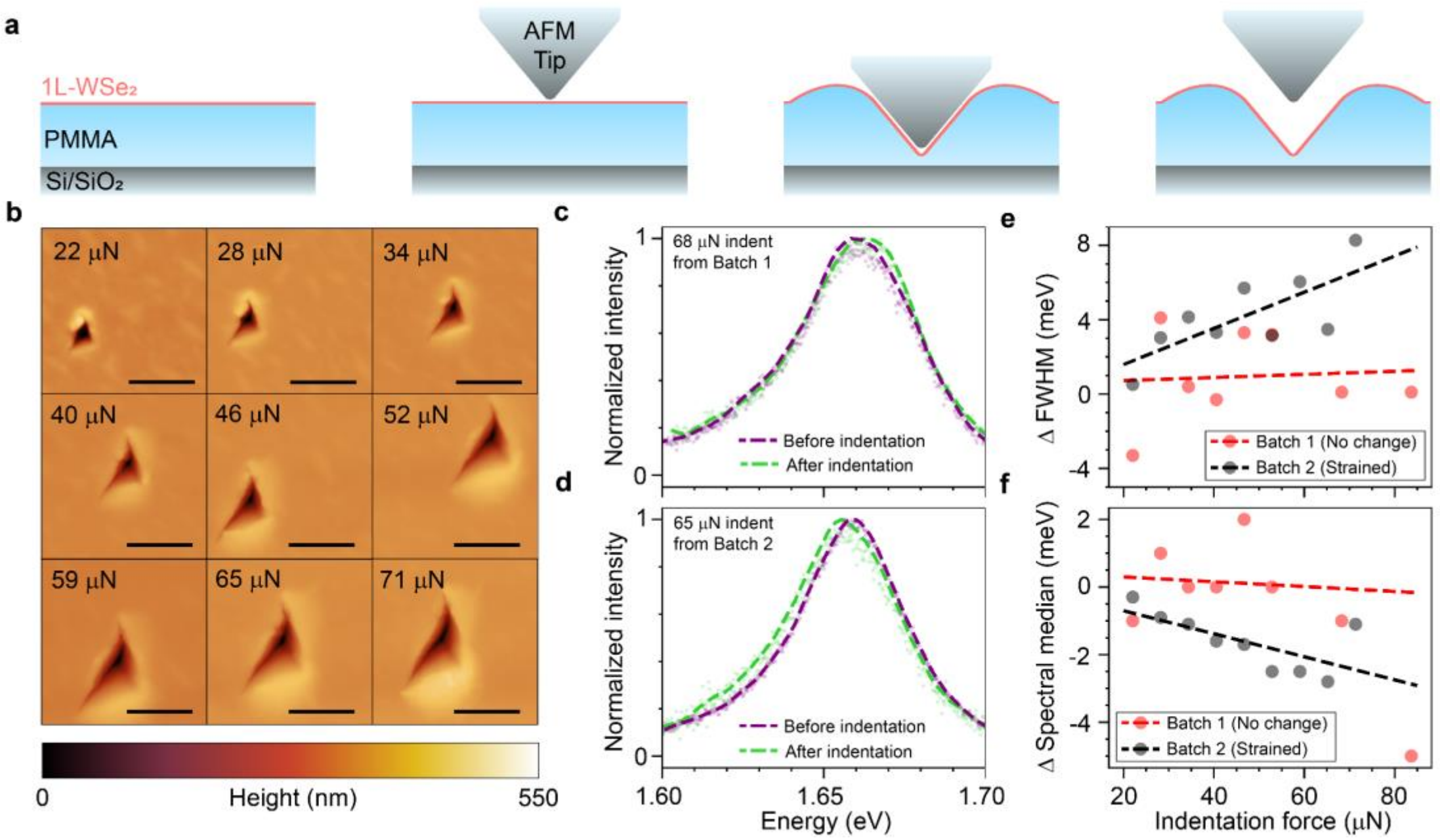

**Figure 1 – The nanoindentation process and room-temperature PL characterization: (a)** A schematic of the indentation process. An AFM tip is positioned over a 1L-$WSe_2$ flake on a PMMA/$SiO_2$ substrate. The substrate is moved into the tip, deforming the $WSe_2$ and PMMA into the shape of the tip. When the substrate is retracted, an indent in the shape of the tip is left. **(b)** High-resolution AFM scans of nine indents from Batch 2 were made with indentation forces ranging from 22 – 71 μN. AFM scale bars are 400 nm. **(c, d)** PL spectra taken before and after indentation from Batch 1 and Batch 2, respectively. **(e)** The change in the full-width half maximum (FWHM) of the PL spectra taken after and before indentation. A positive difference corresponds to an increase in FWHM after indentation. **(f)** The change in PL emission energy, calculated by taking the difference in the PL emission energy before and after indenting. Negative values correspond to a redshift or decrease in emission energy after indenting.

Prior studies of localized strain in 2D semiconductors generated by nanoindentation used blunted probes and $z$-displacements ranging from 200-2000 nm[32]. Using the cantilever force constant, the applied forces in these studies spanned 1-80 μN. In comparison, we use as-supplied commercial Tap300Al-G probes (Figure S1) and $z$-displacements of 1000-3000 nm corresponding to forces from 22-84 μN (see methods and Figure S2). Figure 1b shows the topography of a representative set of indents from our process, demonstrating how the structure of the asymmetric buildup regions, as well as the size and depths of the indents, depend on the indentation force. To probe the strain generated, PL was collected from a 900×900 $nm^2$ region centered around each indentation before and after indentation. The before-and-after comparison of the PL spectra reveals two distinct behaviors. In the first case (Figure 1c), the PL of the indented region is unaffected by the indentation process: both the peak energy and spectral width of the PL are the same before and after the indentation process. The lack of observable change in the PL indicates that the nanoindentation process did not impart a significant amount of strain to the region. In the second case (Figure 1d), the indentation process shifts the PL spectrum to lower energies and increases the spectral linewidth. Such lower energy and broader PL are expected from a region with localized, inhomogeneous tensile strain[8, 19, 22, 37-40], indicating that the indentation process generated a measurable amount of strain. We note that the spatial resolution of these farfield measurements is diffraction-limited, making them unable to resolve the nanoscale strain distributions and leaving ambiguity about how the strain is distributed across the indent, buildup region, and periphery.

For a given "batch" of indents (i.e., a set of nanoindents created with the same AFM probe and identical tip-sample-cantilever geometry), the before-and-after behavior is uniform up to a critical indentation force: either all indents exhibit signatures of localized strain or not. Figures 1e and 1f report how the nanoindentation-induced changes in the energy and width of the PL of 1L-$WSe_2$ depend on the indentation force for both types of batches. Neither the energy nor the linewidth of the PL substantially changes for the set of indents labeled "Batch 1". For this batch, signatures of strain are only observed at the highest indentation force of 84 μN ($z$-displacement of 3000 nm), resulting in an abrupt redshift of ~5 meV. We suspect that this shift at the greatest force is due to a significant structural perturbation beyond indentation, which has been observed in prior studies of indents with lateral dimensions on the length scale of 4 μm[27]. In contrast, for the indents of Batch 2, greater indentation forces generate both larger redshifts and increases in linewidth, indicating force-induced increases of strain and structural inhomogeneity. In particular, the shifts in emission energy are consistent with observations made for similar strain profiles of prior studies[14]. These approximately linear relationships indicate that with the right processing conditions (e.g., AFM probe and tip-sample-cantilever geometry), the indentation force provides a degree of tunability of the localized strain. The different behaviors between the two batches of indents suggest that the indentation-induced strain in the indents of Batch 1 has relaxed, likely due to the tearing of the 1L-$WSe_2$[41], whereas in Batch 2, the 1L-$WSe_2$ remains intact, preserving the indentation-induced strain.

To more directly probe the presence of tearing within the indented material, we developed a process to invert the indents into protruding pillars so that the structure of the 1L-$WSe_2$ within the indent can be assessed using scanning electron microscopy (SEM). For inversion, we follow a template stripping process that is summarized in Figure 2a. In short, a 550 nm-thick gold film is deposited on top of the indented 1L-$WSe_2$/PMMA stack. Care is taken to ensure the evaporated gold is incident parallel to the normal surface to minimize shadow masking and ensure the indents are uniformly filled (see methods and Figure S3). After deposition, a second Si/$SiO_2$ substrate is epoxied to the exposed surface of the gold film. Once cured, the Si/$SiO_2$/epoxy/Au/1L-$WSe_2$/PMMA is stripped from the original substrate, and the residual PMMA is removed by rinsing the substrate with solvent. The gold is in direct contact with the 1L-$WSe_2$, quenching all PL emission in the inverted structure. This inversion process has a high success rate of ~90%, which we suspect is aided by chalcogen vacancies in the TMDs binding with the gold layer[42-45]. With that said, the inversion process also works for other 2D materials ($WS_2$, h-BN) and filling layers (PDMS, $SiO_2$, Ti).

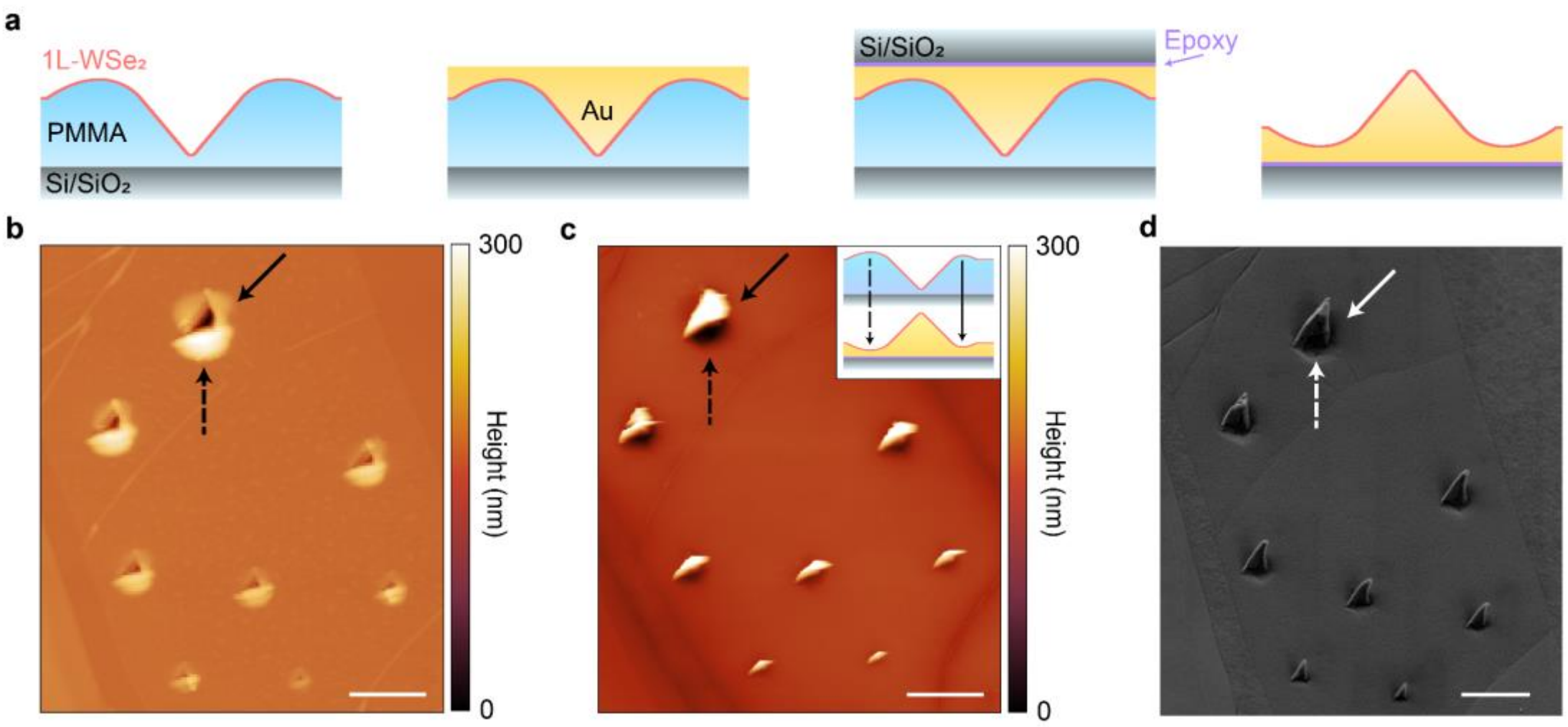

**Figure 2 – Inverting the nanoindents: (a)** A schematic demonstrating the inversion process. Starting with an indented 1L-$WSe_2$/ PMMA/$SiO_2$/Si sample, a ~550 nm thick layer of gold is deposited on the sample, filling the indent with gold, and leaving a gold surface on which a second Si/$SiO_2$ substrate is epoxyed. The PMMA holding the initial Si/$SiO_2$ substrate is dissolved, leaving an inverted indent. **(b, c)** AFM topography maps of Batch 1 indents and the corresponding inverted structure, respectively. The inset in (c) is a schematic demonstrating that the asymmetric PMMA buildup in the indents becomes valleys when inverted. The black (white) dashed and solid lines mark how the larger and smaller buildup regions around the indent correspond to the larger and smaller valleys when inverted. **(d)** SEM of the inverted indents in panel c. The scale bars are 1 μm.

The inverted structure is a coarse replication of the AFM probe that was used for the indentation (see Figure S4). However, the radius of curvature of the apex of the inverted structure is ~2-5× larger than that of the indenting AFM probe. This enlargement is likely caused by both the slipping of the AFM probe during indentation and the difference in elastic modulus between PMMA and 1L-$WSe_2$ (see Figure S5)[46, 47]. Like prior nanoindentation studies[14, 26-34], a bulge is present around the periphery of the indent/inversion (black/white solid and dashed arrows in Figures 2b-d). This bulge arises from the buildup of PMMA as it is pushed away from the indented region but is not uniformly distributed around the indent. Most likely, the tilt of the indenting AFM probe concentrates PMMA on one side of the indent (Figure S6). We note that the asymmetric bulge is greater in the indents of Batch 1 (Figure 2b) than those of Batch 2 (Figure 1b), which is likely due to the differences in probe and indentation geometry. High-resolution AFM micrographs of the indents and inversions confirm the asymmetric bulge. The bulge in the indentation appears as a depression in the inverted structures, demonstrating how the inversion process preserves the structure of the original indents (cf. Figure 2c, inset). Figure 2d shows a SEM micrograph of the inverted structures at a 45° tilt. Within these images, the structure of the 1L-$WSe_2$ is resolvable, confirming that the 1L-$WSe_2$ is preserved during the inversion process.

The tearing of the 1L-$WSe_2$ can be assessed with more detailed SEM imaging (again at a 45° tilt). Figure 3 shows higher-resolution imaging of individual indentations from Batch 1 (Figures 3a) and Batch 2 (Figures 3b) from both sides of the indent. The indents from Batch 1, where the room-temperature spectroscopy indicated the occurrence of strain relaxation, have dark voids that are spanned by filaments of 1L-$WSe_2$ at the depressions that occur at the base of the inverted indents. These features signify tearing of the 1L-$WSe_2$ and generally are only observed on one side of the inverted structure. Further, they occur on the same side for a given batch of indents. Additional examples of torn inversions and their dependence on indentation force, contact angle, and the condition of the AFM probe can be found in Figure S7. In contrast, the indents from Batch 2, where the spectroscopy at room-temperature showed indentation force-dependent strain, are devoid of analogous signatures of tearing. We note that the indents of Batch 1 and Batch 2 were made using nearly the same process. The only variations were the specific probe used and unavoidable slight differences in the specific geometry of how it was mounted in the AFM.

To investigate the tearing further, we used the inverted structures to model the strain in a *hypothetical* 2D material that is distorted to the measured shape (Figure S8)[48]. This analysis finds excessively large strains in the hypothetical structure, which is likely to trigger mechanical failure and tearing if strain relaxation is not possible[47, 49]. When the indents remain intact, we suspect that the strain is relaxed through a combination of slippage of $WSe_2$ and/or shearing of the PMMA. In the case of the torn indents, we hypothesize that there are several factors that contribute to mechanical failure in addition to the large amounts of strain. First, and most importantly, is the use of a dulled AFM probe during indentation, creating a larger and less uniform contact surface. Second is a misalignment in the contact angle used, causing the tip to slide on the $WSe_2$ surface from an insufficient/excessive in-plane motion of the stage. Lastly is the yaw tilt of the probe within the AFM tip mount, causing spatially inhomogeneous strain generation during the indent. Future studies that systematically characterize and control for these factors are needed to determine their relative contributions to the tearing.

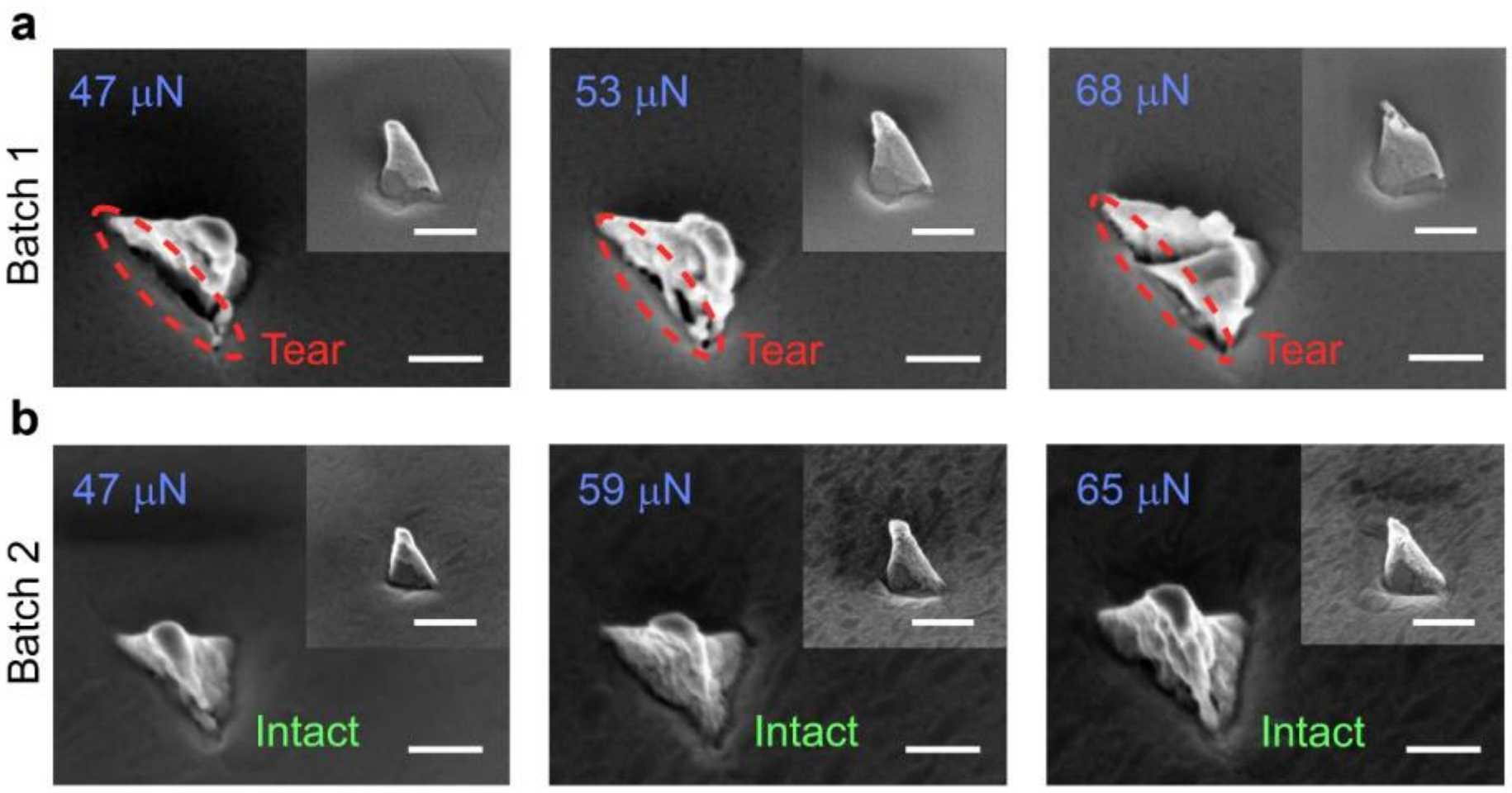

**Figure 3 – SEM of Batch 1 and 2 inverted indents: (a)** SEM images of inverted indents from Batch 1. The red dash ovals mark tears in the 1L-$WSe_2$. This tearing is in sharp contrast to Batch 2, where no tearing is observed in **(b)**. The insets are of the same inverted indents but rotated 180° to show the other side. The blue values in each panel are the indentation forces used to make the indent. Main figure scale bars are 200 nm (insets are 400 nm).

As shown in prior studies, the nanoindentation procedure semi-deterministically positions quantum emitters in 1L-$WSe_2$[14, 26-34]. However, the role of the observed tearing in both the formation and strain-engineering of the quantum emitters has not been established. Addressing this question, Figure 4 shows that quantum light sources emerge in both batches of indents. Localized emission in narrow wavelength bands is observed at the indentation sites for the torn indents of Batch 1 (Figure 4a and inset). However, as contrasted in Figure 4b, the PL of the intact indents is much more uniformly localized to the indentation sites. Detailed imaging of the indents over different emission bands is provided in Figures S9-S10. The spectra of the localized PL at cryogenic temperatures from a representative torn indent in Batch 1 (where the strain has at least partially relaxed) and an intact indent in Batch 2 (where the strain is preserved) are shown in the main panels of Figures 4a and 4b, respectively. Despite the disparity in strain relaxation, narrowband emission from localized states is observed from both types of indentation sites. The quantum nature of the emission from both indentation sites is confirmed with antibunching measurements presented in Figures 4c and 4d. For the torn indentation (Figure 4c), a high single photon purity of 80% is observed, whereas the brightest emitter from the intact indent (Figure 4d) overlaps with a broadband background, reducing the single photon purity to 47% (see Figure S11 for an expanded range of delays that established the background levels). To improve the single-photon sources in 1L-$WSe_2$ nanoindents, the use of low-defect density flux-synthesized $WSe_2$[23, 50] or an additional graphite adlayer[28] has been shown to produce high-purity SPEs. It is well known that localized tensile strain is a critical precursor to the formation of quantum light sources in 1L-$WSe_2$[4, 6, 51, 52], so the emergence of these states in the torn, strain-relaxed indents is not immediately anticipated. Although the strain is relaxed, we suspect that localized strain exists at the tear, providing a secondary route for the formation of quantum emitters[6]. The properties of such emitters (i.e., number of emitters, emitter energy, brightness, etc.) are not expected to depend on the indentation force (beyond achieving mechanical failure) since the strain beyond the tear is relaxed. Several prior studies have reported a similar weak or nonexistent dependence of emitter properties on indentation force[32, 34], indicating that tear-induced formation of quantum emitters could be prevalent in these systems.

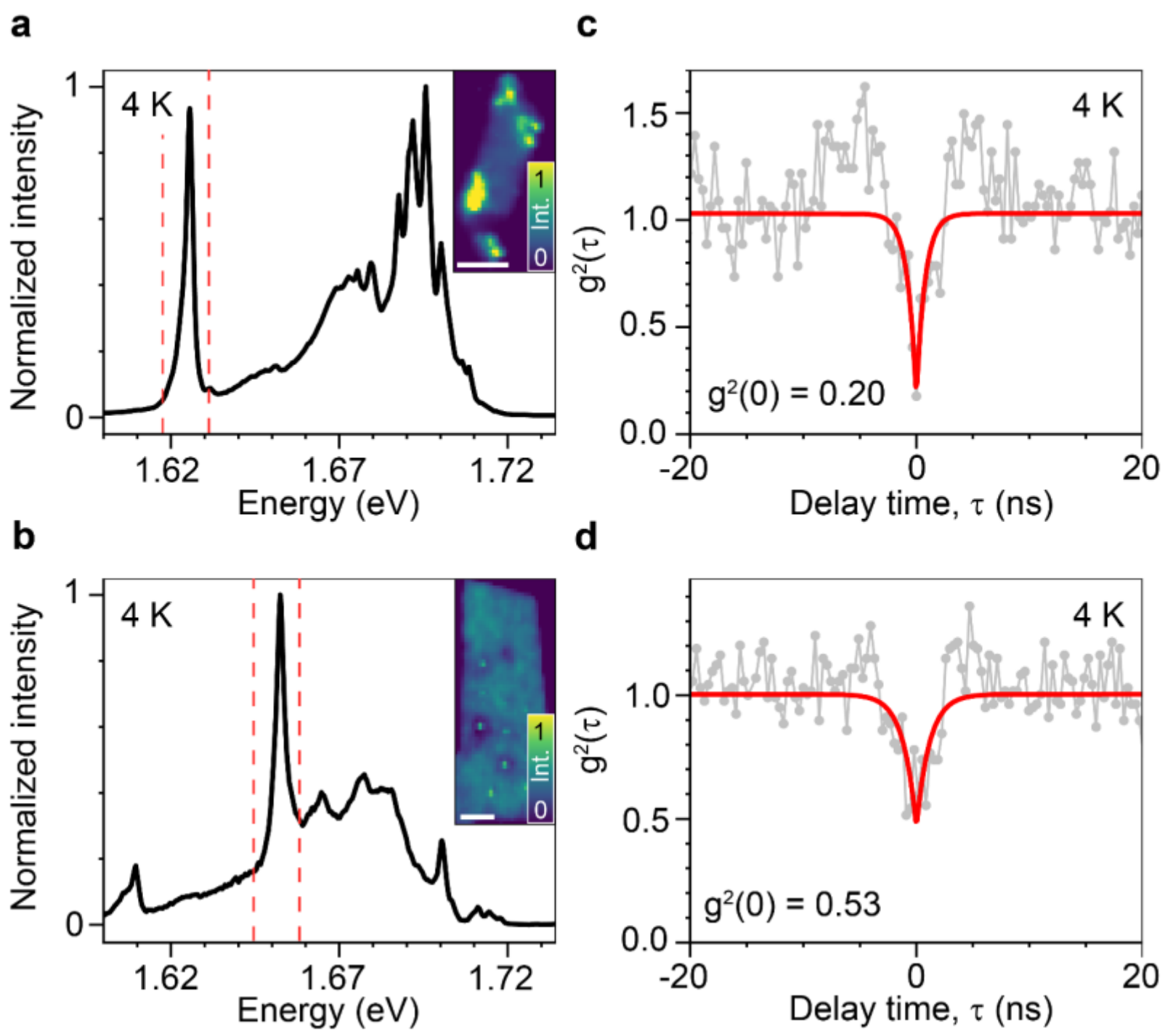

**Figure 4 – Confirmation of single-photon emitters in Batch 1 and 2: (a, b)** PL spectra at 4 K of an indent from Batch 1 and Batch 2, respectively. The inset in (a) is the integrated intensity map of Batch 1 integrated over the range 1.52 – 1.84 eV. The inset in (b) is the integrated intensity map of Batch 2 integrated over the range 1.51 – 1.83 eV. The scale bars of the insets are 5 μm. Both of the insets are on PMMA, so the PL is not quenched. **(c, d)** Second-order correlation function ($g^2(\tau)$) of the emitters marked by red dashed lines in (a) and (b), respectively. Both emitters measured show antibunching behavior with $g^2(0)$ below or near 0.5.

Finally, for both the intact and torn indents, we measured the strain-tunability of the emitter populations and their properties for indentation forces ranging from 22-71 μN (1000-2600 nm *z*-displacement). As shown in the Figure S12, even after verifying the presence of strain in the indents (c.f. Figure 1), clear relationships between the indentation force, emitter energies, and brightnesses were not observed, indicating that the ability to use the indentation force alone to imbue emitters with specific characteristics is limited. However, as shown in Figure 5, the indentation force affects the number and spatial density of emitters that form at an indent. To measure the emitter populations meaningfully, two major challenges exist for obtaining accurate counts. First, a broadband background emission obscures weaker narrowband emitters (Figure 5a). So, first, we counted emitters in the presence of this pronounced background. Then, to suppress this background, we overlaid a batch of intact indents in 1L-$WSe_2$ with a thin layer of graphite (~8 nm), which was previously shown to quench the background and improve the spectral isolation of the localized emitters[28]. AFM characterization reveals that the graphite is suspended over the indent, supported by the surrounding buildup region (see Figure S13). The enhanced PL emission of the localized states over the background is shown in the lower panel of Figure 5a. In addition to a strong suppression of the background, the addition of the graphite activates a larger number of additional emitters that were not resolvable in the original indent. The second challenge to the counting process is the fluctuations of the emitters as exemplified in the temporal evolution of the PL spectrum of a single indentation site (inset, Figure 5a). To reduce the effects of these fluctuations, we identified emissive states in the time-averaged spectra shown in Figures 5a and then confirmed that each line displayed independent fluctuations (i.e., spectral diffusion/wandering) using PL time series such as the inset in Figure 5a[53]. For instance, the emission peaks encircled in the black-dashed line in Figure 5a are examples of single states that are fluctuating between two dominant energies. Individual PL spectra used for emitter counting can be found in Figures S14-S16.

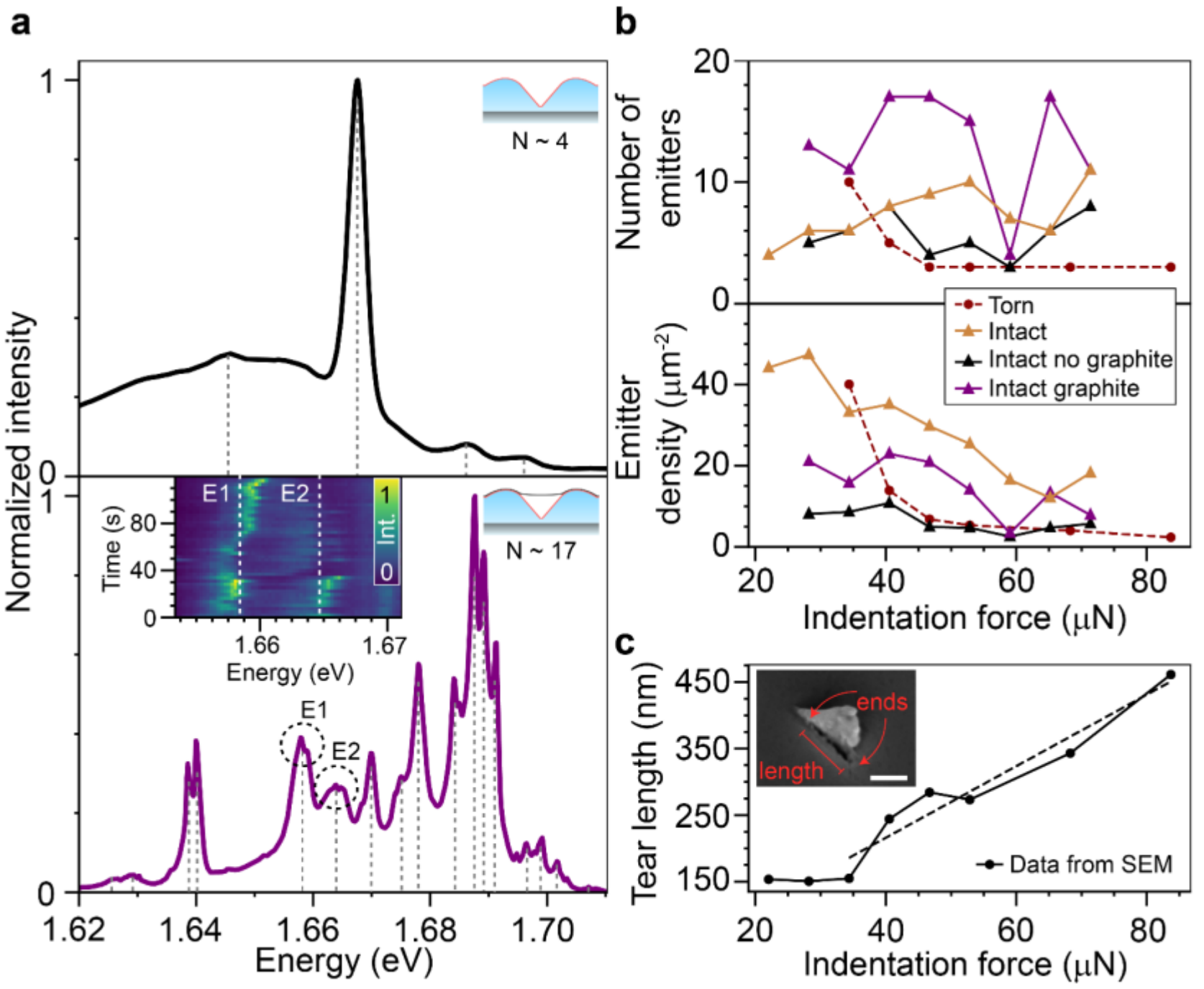

**Figure 5 – Emitter density and emitter count dependence on indentation force: (a)** Time-averaged 4 K PL spectra from an intact indent without graphite (top) and an intact indent with graphite (bottom). N is the number of emitters counted from the average spectra. The schematics in (a) represent the indents with or without graphite. The inset in (a, bottom) is the time series of the emitters circled in black and marked E1 and E2. The corresponding emitters in the time series are marked with white dashed lines. **(b)** Emitter number and emitter density with respect to indent area as a function of indentation force. In (b), the red circles correspond to Batch 1 (torn), and the yellow triangles correspond to Batch 2 (intact). The Batch 3 indents without graphite are marked by the black triangles, and the indents with graphite are represented by the purple triangles. **(c)** The tear length of each indent in Batch 1 as a function of indentation force. The tear lengths were measured from SEM images. The dashed black line is a guide to the eye. An example of how the tear was measured is shown in the SEM inset. SEM image scale bar is 200 nm.

Following the above counting protocol, Figure 5b shows how the number of emitters depends on the indentation force for torn indents, intact indents, and intact indents overlaid with graphene. The probed indentation forces range from 22-84 μN (1000-3000 nm $z$-displacement). For Batches 1 (torn) and 3 (intact without graphite and intact with graphite), the smallest forces used did not generate localized emitters, and we excluded those points from the plot. For all cases where the indents are intact, the absolute number of emitters weakly depends on the indentation force for the smallest forces, which is most evident in the intact indents of Batch 2 (intact). Beyond 41 μN (~1600 nm $z$-displacement), the number of emitters appears to be independent of the indentation force. This general behavior was observed in two independent batches of indents. We note the indent made with 59 μN (2200 nm $z$-displacement) in Batch 3 is a significant outlier. The overall PL intensity at this site was also vastly reduced, indicating that an issue likely occurred during the indentation process for this site.

The trend for the torn indents is dramatically different: as the indentation force increases, the number of emitters *decreases* with a rate that drastically slows beyond ~50 μN. While the number of emitters reveals a clear discrepancy between intact and torn indents, a strong dependence on the force is not observed for the intact indents. In contrast, Figure 5b shows that the *areal density* of the localized emitters has a stronger dependence on the indentation force. The surface area, which includes the indented and buildup regions, is determined with high-resolution AFM imaging. For intact indents, the emitter density decreases linearly with increasing force, indicating that either quantum emitter formation favors a specific feature of the indent, such as its apex or surrounding buildup region. For instance, recent studies have shown the importance of nanoscale wrinkles in the formation of quantum emitter states in 1L-$WSe_2$[18, 50, 54]. A potential explanation is that such wrinkles are suppressed within the indent and confined to the surrounding regions, which anecdotally agrees with prior super-resolution measurements[26]. Empirically, the total area and the area of buildup regions increase linearly with indentation force (see Figure S17). In contrast, the surface area of the indent increases quadratically. Thus, localization of the emitters to the buildup region would lead to the observed decrease in emitter density. We also note that a similar effect would be observed if the particular feature corresponds to a specific range of tensile strains within the indent where emitter formation can occur. The possibility of such an upper limit on strain for emitter formation has not been explored, to the best of our knowledge.

The more dramatic decrease in both the absolute number of emitters as well as the emitter density for the torn indents supports our hypothesis that the emitters are localized to the tear. As shown in Figure 5c, the length of the tear is constant up to 34 μN and then increases linearly with indentation force (see Figure S18 for SEM images). At 34 μN, an abrupt increase in the number of emitters occurs, which then rapidly decreases with greater forces until a constant number of three emitters. We believe that this behavior indicates that the emitters are localized to specific regions of the tear. If this is not the case and the emitters are uniformly distributed along the length of the tear, we would expect a linearly decreasing trend as seen in the intact indents. We suspect that the 34 μN threshold marks a critical threshold for strain relaxation in the $1L\text{-}WSe_2$. Along these lines, we interpret the rapid decrease of the number of emitters between 34-48 μN to increasing relaxation of the strain by the tear until maximum relaxation beyond 48 μN. At the point of maximum relaxation, we only observe three emitters per site, regardless of the force used.

For the intact indents, whether the localized emitters are predominantly in the indent itself or in the peripheral buildup region around the indent remains unresolved. While, as noted above, the decrease in the emitter density could be explained by localization to the buildup region, a similar trend would be observed if emitter formation requires tensile strains in a closed range of values (i.e., requires a strain between a minimum and maximum value). Further, we observe that overlaying the indents with graphite dramatically enhances the formation of the SPE states, which is not immediately expected if the emitters are preferentially located in the buildup region: any emitters in the buildup region should be preferentially quenched due to their proximity to the graphite. Further exploration of this issue motivates future studies that, for instance, leverage a dielectric material for the inversion process so that emitters are not quenched and can be localized using nano-optical or super-resolution microscopy techniques[18, 25, 50, 55-57]. More quantitative structural models of the strain in both the torn and intact indents that are able to account for effects such as slippage of the $WSe_2$ and/or shearing of the PMMA would shed additional light on this open question.

**Conclusions**

The ability to deterministically position emitters with localized tensile strain is a major advantage of the single-photon emitters $1L\text{-}WSe_2$. Compared to other approaches where the strain is deterministically generated by dry-transferring the $1L\text{-}WSe_2$ onto nanofabricated structures such as nanopillars[22, 26, 58], nanocones[18, 21, 50], and trenches/holes[10, 59] nanoindentation provides a more reproducible deformation of the 2D material (especially considering wrinkling). However, emitters from nanoindents typically have inferior single-photon purity, brightness, polarization control, and photon indistinguishability compared to these other platforms. These challenges are evident in our study here: strong background emission (Figure 4) reduces the single-photon purity, and the temporal fluctuations (Figure 5) are detrimental to photon indistinguishability. For the indents, substantial improvements in purity have been achieved in prior work by either adding a top layer of graphite[28] (as repeated here; Figure 5) or electrostatic gating[33]. Little work has been done to optimize the spectral stability of indent-derived emitters, including exploration of resonant and pulsed excitation, which has improved the emitter quality in both non-indented 1L-$WSe_2$ [60, 61] and 2L-$WSe_2$ [62] systems. Further, integration of a thin layer of hBN in the indentation/inversion process to further stabilize the emitters as recently shown for pillar-induced SPEs[63] has yet to be explored, likely due to the necessity of thin layers of hBN that remain mechanically compliant. Despite these current limitations, nanoindents are advantageous for investigating structure-property relationships of single-photon emitters in strained 1L-$WSe_2$ because the deformation of the 2D material is highly controllable and reproducible. As only the deformation process is changing, the structure-property relationships uncovered by the indents should be deeply relevant to strain-engineering emitters using other stressor architectures.

Building on the controllability of the nanoindentation process, our study provides a detailed characterization of the structure of indented 1L-$WSe_2$. Using a process to invert the indents, we assess the physical structure of the indented 1L-$WSe_2$ and confidently differentiate between indents where the 2D semiconductor is torn from those where it remains intact. For the intact indents, clear signatures of strain are identified in room-temperature PL characterization: the indentation process results in a localized red-shift and broadening of the PL. The magnitudes of the red shift and broadening increase with increasing indentation force, indicating that larger indentation forces generate greater amounts of localized strain. In contrast, in the idents where the 2D semiconductor is torn, the tear sufficiently relaxes the localized strain such that neither a shift nor broadening of the PL spectrum can be resolved. Remarkably, when focusing on the intact indents by eliminating cases where the 2D semiconductor has torn, the indentation force does not systematically alter the emitter energies nor the emitter intensities. However, we find that the larger localized strains from the greater indentation forces reduce the density of the localized emitters within an indent. This observation provides strong evidence that the emitters are localized to a limited region of the indent. Possible explanations include that the emitters only form over a limited range of tensile strain or are localized to the surrounding buildup region as was concluded from localization microscopy [26]. For the indents where tearing occurs, the dependence of the number of emitters on the indentation force is dramatically different, indicating that the localized emitters are isolated to specific regions of the indent, likely the tears. These measurements provide critical insight into the mechanisms and limitations of deterministic strain localization and engineering of SPEs in 2D semiconductors, both of which are critical for their development as light sources for quantum technologies. In the future, inverting an indented hBN/1L-$WSe_2$ heterostructure could mitigate the quenching of emission after inversion, opening a new set of possible nano-optical and super-resolution measurements that could provide greater insight into the strain profile of the indented structure and the nanoscale locations of emitters. Improved quantitative understanding of these processes will enable refined approaches that further improve the reliability of localizing high-quality, background-free emitters to specific positions for quantum photonic applications.

**Supporting Information:** additional structural and optical characterization of the indents and inversions, thin plate modeling of the strain, fabrication details, extended-delay second-order correlation measurements of the SPEs, strain-independent emitter properties.

## Methods

### Sample fabrication

**Exfoliating $WSe_2$ onto PMMA:** Polymethyl methacrylate (PMMA) layers on Si/$SiO_2$ were made by spinning 950K A4 PMMA at 1500 RPM, followed by an 80 °C bake for 45 mins. $WSe_2$ flakes were then exfoliated onto PMMA using an adapted Scotch tape method[64]. First, $WSe_2$ was exfoliated onto X4 polydimethylsiloxane (PDMS) that was treated in a light UV ozone for 2 mins, and then the $WSe_2$/PDMS was used to exfoliate onto PMMA that was treated in the same UV ozone. Here, PDMS was used to exfoliate onto PMMA rather than typical tapes due to the reduced residue we observed.

**Generation of nanoindents:** To generate nanoindents in 1L-$WSe_2$, we followed the initial protocol set by Rosenberger *et al.*[32]. We used an AIST-NT SmartSPM HE002 AFM from HORIBA Scientific. In the AFM software, we used the nanoindentation application in the software in "Zsweep" mode, which allows for control of the Z-axis piezo motor of the AFM sample stage. When the location of an indent was defined, the desired indentation force was set by inputting the $z$-displacement, which is the distance the AFM stage moved into the AFM tip. The $z$-displacement ranged from 1000-3000 nm, corresponding to 22-84 μN in indentation force. At these indentation forces, the AFM tip slips, meaning the tip does not stay in the defined location throughout the indentation process. To minimize the slipping, we introduced a lateral movement of the sample stage along the y-axis to keep the tip in the same initial position during the indentation process (see Figure S5).

**Inversion of nanoindents:** To invert a batch of nanoindented 2D materials on PMMA, samples are placed in an electron-beam evaporator directly above the crucible to ensure normal incidence during evaporation. For Au inversion, a layer of Au 150 – 200 nm thicker than the deepest indent depth of a given batch was deposited at 2 Å/s (550 nm in total). Due to the PMMA surface, certain compact evaporators may have issues of polymerized PMMA or substantial film stress caused by excessive electron and charged ion irradiation. This issue can be solved with added magnets and shields[65]. After evaporation, a thin layer of epoxy is spun onto the exposed Au surface at 6000 RPM, and a second Si/$SiO_2$ substrate is attached. Once cured, a razor blade stuck into the corners of the sample stack was used to separate the two substrates. The 2D material/Au/epoxy/$SiO_2$/Si was then quickly rinsed in acetone to remove residual PMMA residue, subsequently cleaned with IPA, and dried with $N_2$.

**Wet transfer of thin graphite:** To transfer graphite layers onto $WSe_2$ indents, graphite was first exfoliated onto a Si/$SiO_2$ substrate using the Scotch tape method[64]. Afterwards, flakes with thicknesses ranging from 6-10 nm were identified using AFM, and a thin layer of 10:1 Sylgard 184 PDMS was spun on at 4000 RPM. After curing in a 100 °C oven for ~60 mins, the chip was submerged in a 1M solution of NaOH in DIW heated to 90 °C to lightly etch the $SiO_2$ surface and encourage liftoff of the graphite/PDMS stamp[66, 67]. Once bubbles formed on the perimeter of the chip (~15 mins), the film was separated from the Si/$SiO_2$, sometimes requiring manual assistance using a pair of tweezers. To remove excess NaOH residue, the film was left in a DIW bath for ~12 hours. Afterwards, the film was retrieved from the DIW, dried with Nitrogen, and transferred over the $WSe_2$ indents using the viscoelastic stamping method[68, 69].

**Optical measurements**

**Room-temperature PL:** The room-temperature PL measurements were conducted in ambient atmosphere using a commercially built microscope (HORIBA Scientific TRIOS system). We coupled a 532 nm CW diode laser in the microscope and focused the laser onto the sample using a Nikon LU Plan Fluor 100X (0.9 NA) objective. In this system, the sample is scanned using an XY piezomotor, and this scanning confocal microscope has a spatial resolution of ~630 nm. The room-temperature PL spectra were recorded using a grating spectrometer (Andor Kymera) and a CCD camera (Andor iXon Ultra 888 EMCCD).

**Cryogenic PL:** Cryogenic PL measurements were taken inside a liquid helium-cooled Montana Instruments s50 Cryostation (4 K) equipped with optical access. 532 nm CW laser light was focused on the sample with a Nikon infinity corrected S Plan Fluor 40X (0.6 NA) objective equipped with a correction collar to compensate for a thin glass window above the sample. This home-built scanning confocal microscope has a spatial resolution of ~1 μm. PL spectra were then recorded using a grating spectrometer (Horiba iHR320) and a CCD camera (Andor idus 416). Single-photon antibunching measurements were conducted by isolating a narrow band of quantum emission using the exit slit of the spectrometer as a monochromator. The narrowband emission was sent into a free-space Hanbury-Brown and Twiss (HBT) interferometer. The single-photon avalanche photodiodes that make up the arms of the HBT were connected to a PicoQuant HydraHarp 400 that acted as our time tagging module. To normalize the antibunching data, we averaged the first 250 data points, located at ~100 ns of delay time, and divided this average value across every histogram bin.

**Acknowledgements:** We acknowledge the MonArk NSF Quantum Foundry, supported by the National Science Foundation Q-AMASE-i program under NSF award No. DMR-1906383. This work was performed in part at the Montana Nanotechnology Facility, SCR_026333, a member of the National Nanotechnology Coordinated Infrastructure (NNCI), which is supported by the National Science Foundation (Grant# ECCS-2025391). Work at the Molecular Foundry was supported by the Office of Science, Office of Basic Energy Sciences, of the U.S. Department of Energy under Contract No. DE-AC02-05CH11231.

**References**

(1) Montblanch, A. R.; Barbone, M.; Aharonovich, I.; Atature, M.; Ferrari, A. C. Layered materials as a platform for quantum technologies. *Nat Nanotechnol* **2023**, *18* (6), 555-571.

(2) Yu, Y.; Seo, I. C.; Luo, M.; Lu, K.; Son, B.; Tan, J. K.; Nam, D. Tunable single-photon emitters in 2D materials. *Nanophotonics* **2024**, *13* (19), 3615-3629.

(3) Sortino, L.; Zotev, P. G.; Phillips, C. L.; Brash, A. J.; Cambiasso, J.; Marensi, E.; Fox, A. M.; Maier, S. A.; Sapienza, R.; Tartakovskii, A. I. Bright single photon emitters with enhanced quantum efficiency in a two-dimensional semiconductor coupled with dielectric nano-antennas. *Nat Commun* **2021**, *12* (1), 6063.

(4) Luo, Y.; Shepard, G. D.; Ardelean, J. V.; Rhodes, D. A.; Kim, B.; Barmak, K.; Hone, J. C.; Strauf, S. Deterministic coupling of site-controlled quantum emitters in monolayer $WSe_2$ to plasmonic nanocavities. *Nat Nanotechnol* **2018**, *13* (12), 1137-1142.

(5) Gao, T.; von Helversen, M.; Antón-Solanas, C.; Schneider, C.; Heindel, T. Atomically-thin single-photon sources for quantum communication. *npj 2D Materials and Applications* **2023**, *7* (4).

(6) Tonndorf, P.; Schmidt, R.; Schneider, R.; Kern, J.; Buscema, M.; Steele, G. A.; Castellanos-Gomez, A.; van der Zant, H. S. J.; Michaelis de Vasconcellos, S.; Bratschitsch, R. Single-photon emission from localized excitons in an atomically thin semiconductor. *Optica* **2015**, *2* (4).

(7) Srivastava, A.; Sidler, M.; Allain, A. V.; Lembke, D. S.; Kis, A.; Imamoglu, A. Optically active quantum dots in monolayer $WSe_2$. *Nat Nanotechnol* **2015**, *10* (6), 491-496.

(8) Li, H.; Contryman, A. W.; Qian, X.; Ardakani, S. M.; Gong, Y.; Wang, X.; Weisse, J. M.; Lee, C. H.; Zhao, J.; Ajayan, P. M.; et al. Optoelectronic crystal of artificial atoms in strain-textured molybdenum disulphide. *Nat Commun* **2015**, *6* (1), 7381.

(9) Koperski, M.; Nogajewski, K.; Arora, A.; Cherkez, V.; Mallet, P.; Veuillen, J. Y.; Marcus, J.; Kossacki, P.; Potemski, M. Single photon emitters in exfoliated $WSe_2$ structures. *Nat Nanotechnol* **2015**, *10* (6), 503-506.

(10) Kumar, S.; Kaczmarczyk, A.; Gerardot, B. D. Strain-Induced Spatial and Spectral Isolation of Quantum Emitters in Mono- and Bilayer $WSe_2$. *Nano Lett* **2015**, *15* (11), 7567-7573.

(11) Moody, G.; Kavir Dass, C.; Hao, K.; Chen, C. H.; Li, L. J.; Singh, A.; Tran, K.; Clark, G.; Xu, X.; Berghauser, G.; et al. Intrinsic homogeneous linewidth and broadening mechanisms of excitons in monolayer transition metal dichalcogenides. *Nat Commun* **2015**, *6* (1), 8315.

(12) He, Y. M.; Clark, G.; Schaibley, J. R.; He, Y.; Chen, M. C.; Wei, Y. J.; Ding, X.; Zhang, Q.; Yao, W.; Xu, X.; et al. Single quantum emitters in monolayer semiconductors. *Nat Nanotechnol* **2015**, *10* (6), 497-502.

(13) Chakraborty, C.; Kinnischtzke, L.; Goodfellow, K. M.; Beams, R.; Vamivakas, A. N. Voltage-controlled quantum light from an atomically thin semiconductor. *Nat Nanotechnol* **2015**, *10* (6), 507-511.

(14) So, J. P.; Kim, H. R.; Baek, H.; Jeong, K. Y.; Lee, H. C.; Huh, W.; Kim, Y. S.; Watanabe, K.; Taniguchi, T.; Kim, J.; et al. Electrically driven strain-induced deterministic single-photon emitters in a van der Waals heterostructure. *Sci Adv* **2021**, *7* (43), eabj3176.

(15) Palacios-Berraquero, C.; Barbone, M.; Kara, D. M.; Chen, X.; Goykhman, I.; Yoon, D.; Ott, A. K.; Beitner, J.; Watanabe, K.; Taniguchi, T.; et al. Atomically thin quantum light-emitting diodes. *Nat Commun* **2016**, *7* (1), 12978.

(16) Schwarz, S.; Kozikov, A.; Withers, F.; Maguire, J. K.; Foster, A. P.; Dufferwiel, S.; Hague, L.; Makhonin, M. N.; Wilson, L. R.; Geim, A. K.; et al. Electrically pumped single-defect light emitters in $WSe_2$. *2D Materials* **2016**, *3* (2).

(17) Guo, S.; Germanis, S.; Taniguchi, T.; Watanabe, K.; Withers, F.; Luxmoore, I. J. Electrically Driven Site-Controlled Single Photon Source. *ACS Photonics* **2023**, *10* (8), 2549-2555.

(18) Yanev, E. S.; Darlington, T. P.; Ladyzhets, S. A.; Strasbourg, M. C.; Trovatello, C.; Liu, S.; Rhodes, D. A.; Hall, K.; Sinha, A.; Borys, N. J.; et al. Programmable nanowrinkle-induced room-temperature exciton localization in monolayer $WSe_2$. *Nat Commun* **2024**, *15* (1), 1543.

(19) Branny, A.; Kumar, S.; Proux, R.; Gerardot, B. D. Deterministic strain-induced arrays of quantum emitters in a two-dimensional semiconductor. *Nat Commun* **2017**, *8* (1), 15053.

(20) Azzam, S. I.; Parto, K.; Moody, G. Purcell enhancement and polarization control of single-photon emitters in monolayer $WSe_2$ using dielectric nanoantennas. *Nanophotonics* **2023**, *12* (3), 477-484.

(21) Kim, H.; Moon, J. S.; Noh, G.; Lee, J.; Kim, J. H. Position and Frequency Control of Strain-Induced Quantum Emitters in $WSe_2$ Monolayers. *Nano Lett* **2019**, *19* (10), 7534-7539.

(22) Parto, K.; Azzam, S. I.; Banerjee, K.; Moody, G. Defect and strain engineering of monolayer $WSe_2$ enables site-controlled single-photon emission up to 150 K. *Nat Commun* **2021**, *12* (1), 3585.

(23) Luo, Y.; Liu, N.; Li, X.; Hone, J. C.; Strauf, S. Single photon emission in $WSe_2$ up to 160 K by quantum yield control. *2D Materials* **2019**, *6* (3).

(24) Shepard, G. D.; Ajayi, O. A.; Li, X.; Zhu, X. Y.; Hone, J.; Strauf, S. Nanobubble induced formation of quantum emitters in monolayer semiconductors. *2D Materials* **2017**, *4* (2).

(25) Darlington, T. P.; Carmesin, C.; Florian, M.; Yanev, E.; Ajayi, O.; Ardelean, J.; Rhodes, D. A.; Ghiotto, A.; Krayev, A.; Watanabe, K.; et al. Imaging strain-localized excitons in nanoscale bubbles of monolayer $WSe_2$ at room temperature. *Nat Nanotechnol* **2020**, *15* (10), 854-860.

(26) Abramov, A. N.; Chestnov, I. Y.; Alimova, E. S.; Ivanova, T.; Mukhin, I. S.; Krizhanovskii, D. N.; Shelykh, I. A.; Iorsh, I. V.; Kravtsov, V. Photoluminescence imaging of single photon emitters within nanoscale strain profiles in monolayer $WSe_2$. *Nat Commun* **2023**, *14* (1), 5737.

(27) Chang, S.; Yan, Y.; Geng, Y. Local Nanostrain Engineering of Monolayer $MoS_2$ Using Atomic Force Microscopy-Based Thermomechanical Nanoindentation. *Nano Lett* **2023**, *23* (20), 9219-9226.

(28) Chuang, H. J.; Stevens, C. E.; Rosenberger, M. R.; Lee, S. J.; McCreary, K. M.; Hendrickson, J. R.; Jonker, B. T. Enhancing Single Photon Emission Purity via Design of van der Waals Heterostructures. *Nano Lett* **2024**, *24* (18), 5529-5535.

(29) Cunningham, P. D.; Proscia, N. V.; LaGasse, S. W.; O'Hara, D. J.; McCreary, K. M.; Chuang, H.-J.; Povolotskyi, M.; Vurgaftman, I.; Jonker, B. T.; Simpkins, B. S. Site-Specific Exciton–Plasmon Coupling in Nanoindented $WSe_2$. *ACS Photonics* **2024**, *11* (8), 3250-3258.

(30) Lee, S.-J.; Chuang, H.-J.; Yeats, A. L.; McCreary, K. M.; O'Hara, D. J.; Jonker, B. T. Ferroelectric Modulation of Quantum Emitters in Monolayer $WS_2$. *ACS Nano* **2024**, *18* (36), 25349-25358.

(31) Li, X.; Jones, A. C.; Choi, J.; Zhao, H.; Chandrasekaran, V.; Pettes, M. T.; Piryatinski, A.; Tschudin, M. A.; Reiser, P.; Broadway, D. A.; et al. Proximity-induced chiral quantum light generation in strain-engineered $WSe_2/NiPS_3$ heterostructures. *Nat Mater* **2023**, *22* (11), 1311-1316.

(32) Rosenberger, M. R.; Dass, C. K.; Chuang, H. J.; Sivaram, S. V.; McCreary, K. M.; Hendrickson, J. R.; Jonker, B. T. Quantum Calligraphy: Writing Single-Photon Emitters in a Two-Dimensional Materials Platform. *ACS Nano* **2019**, *13* (1), 904-912.

(33) Stevens, C. E.; Chuang, H. J.; Rosenberger, M. R.; McCreary, K. M.; Dass, C. K.; Jonker, B. T.; Hendrickson, J. R. Enhancing the Purity of Deterministically Placed Quantum Emitters in Monolayer $WSe_2$. *ACS Nano* **2022**, *16* (12), 20956-20963.

(34) Yücel, O.; Yagodkin, D.; Kirchhof, J. N.; Yu, Y.; Kumar, A. M.; Dewambrechies, A.; Kovalchuk, S.; Bolotin, K. I. Strain activation of localized states in $WSe_2$. *2D Materials* **2025**, *12* (3).

(35) Peng, L.; Chan, H.; Choo, P.; Odom, T. W.; Sankaranarayanan, S.; Ma, X. Creation of Single-Photon Emitters in $WSe_2$ Monolayers Using Nanometer-Sized Gold Tips. *Nano Lett* **2020**, *20* (8), 5866-5872.

(36) Lee, S. J.; Chuang, H. J.; McCreary, K. M.; Noyan, M. A.; Jonker, B. T. Voltage-Induced Degradation for Enhanced Purity and Reproducibility of Quantum Emission in Monolayer 2D Materials. *ACS Nano* **2025**, *19* (38), 33981-33990.

(37) Aslan, B.; Deng, M.; Heinz, T. F. Strain tuning of excitons in monolayer $WSe_2$. *Physical Review B* **2018**, *98* (11).

(38) Conley, H. J.; Wang, B.; Ziegler, J. I.; Haglund, R. F., Jr.; Pantelides, S. T.; Bolotin, K. I. Bandgap engineering of strained monolayer and bilayer $MoS_2$. *Nano Lett* **2013**, *13* (8), 3626-3630.

(39) Desai, S. B.; Seol, G.; Kang, J. S.; Fang, H.; Battaglia, C.; Kapadia, R.; Ager, J. W.; Guo, J.; Javey, A. Strain-induced indirect to direct bandgap transition in multilayer $WSe_2$. *Nano Lett* **2014**, *14* (8), 4592-4597.

(40) Hsu, W. T.; Lu, L. S.; Wang, D.; Huang, J. K.; Li, M. Y.; Chang, T. R.; Chou, Y. C.; Juang, Z. Y.; Jeng, H. T.; Li, L. J.; et al. Evidence of indirect gap in monolayer $WSe_2$. *Nat Commun* **2017**, *8* (1), 929.

(41) Taghinejad, H.; Eftekhar, A. A.; Campbell, P. M.; Beatty, B.; Taghinejad, M.; Zhou, Y.; Perini, C. J.; Moradinejad, H.; Henderson, W. E.; Woods, E. V.; et al. Strain relaxation via formation of cracks in compositionally modulated two-dimensional semiconductor alloys. *npj 2D Materials and Applications* **2018**, *2* (10).

(42) Panasci, S. E.; Schiliro, E.; Greco, G.; Cannas, M.; Gelardi, F. M.; Agnello, S.; Roccaforte, F.; Giannazzo, F. Strain, Doping, and Electronic Transport of Large Area Monolayer $MoS_2$ Exfoliated on Gold and Transferred to an Insulating Substrate. *ACS Appl Mater Interfaces* **2021**, *13* (26), 31248-31259.

(43) Chen, S.; Li, B.; Dai, C.; Zhu, L.; Shen, Y.; Liu, F.; Deng, S.; Ming, F. Controlling Gold-Assisted Exfoliation of Large-Area $MoS_2$ Monolayers with External Pressure. *Nanomaterials (Basel)* **2024**, *14* (17).

(44) Smith, T. The hydrophilic nature of a clean gold surface. *Journal of Colloid and Interface Science* **1980**, *75* (1), 51-55.

(45) Velicky, M.; Donnelly, G. E.; Hendren, W. R.; McFarland, S.; Scullion, D.; DeBenedetti, W. J. I.; Correa, G. C.; Han, Y.; Wain, A. J.; Hines, M. A.; et al. Mechanism of Gold-Assisted Exfoliation of Centimeter-Sized Transition-Metal Dichalcogenide Monolayers. *ACS Nano* **2018**, *12* (10), 10463-10472.

(46) Stafford, C. M.; Harrison, C.; Beers, K. L.; Karim, A.; Amis, E. J.; VanLandingham, M. R.; Kim, H. C.; Volksen, W.; Miller, R. D.; Simonyi, E. E. A buckling-based metrology for measuring the elastic moduli of polymeric thin films. *Nat Mater* **2004**, *3* (8), 545-550.

(47) Falin, A.; Holwill, M.; Lv, H.; Gan, W.; Cheng, J.; Zhang, R.; Qian, D.; Barnett, M. R.; Santos, E. J. G.; Novoselov, K. S.; et al. Mechanical Properties of Atomically Thin Tungsten Dichalcogenides: $WS_2$, $WSe_2$, and $WTe_2$. *ACS Nano* **2021**, *15* (2), 2600-2610.

(48) Darlington, T. P.; Krayev, A.; Venkatesh, V.; Saxena, R.; Kysar, J. W.; Borys, N. J.; Jariwala, D.; Schuck, P. J. Facile and quantitative estimation of strain in nanobubbles with arbitrary symmetry in 2D semiconductors verified using hyperspectral nano-optical imaging. *J Chem Phys* **2020**, *153* (2), 024702.

(49) Ben-Smith, A.; Choi, S. H.; Boandoh, S.; Lee, B. H.; Vu, D. A.; Nguyen, H. T. T.; Adofo, L. A.; Jin, J. W.; Kim, S. M.; Lee, Y. H.; et al. Photo-oxidative Crack Propagation in Transition Metal Dichalcogenides. *ACS Nano* **2024**, *18* (4), 3125-3133.

(50) Strasbourg, M. C.; Yanev, E. S.; Darlington, T. P.; Faagau, K.; Holtzman, L. N.; Barmak, K.; Hone, J. C.; Schuck, P. J.; Borys, N. J. Characterization of quantum dot-like emitters in programmable arrays of nanowrinkles of 1L-$WSe_2$. *Journal of Applied Physics* **2024**, *136* (4).

(51) Dang, J.; Sun, S.; Xie, X.; Yu, Y.; Peng, K.; Qian, C.; Wu, S.; Song, F.; Yang, J.; Xiao, S.; et al. Identifying defect-related quantum emitters in monolayer $WSe_2$. *npj 2D Materials and Applications* **2020**, *4* (2).

(52) Linhart, L.; Paur, M.; Smejkal, V.; Burgdorfer, J.; Mueller, T.; Libisch, F. Localized Intervalley Defect Excitons as Single-Photon Emitters in $WSe_2$. *Phys Rev Lett* **2019**, *123* (14), 146401.

(53) Schneider, L. M.; Lippert, S.; Kuhnert, J.; Ajayi, O.; Renaud, D.; Firoozabadi, S.; Ngo, Q.; Guo, R.; Kim, Y. D.; Heimbrodt, W.; et al. The influence of the environment on monolayer tungsten diselenide photoluminescence. *Nano-Structures & Nano-Objects* **2018**, *15*, 84-97.

(54) Daveau, R. S.; Vandekerckhove, T.; Mukherjee, A.; Wang, Z.; Shan, J.; Mak, K. F.; Vamivakas, A. N.; Fuchs, G. D. Spectral and spatial isolation of single tungsten diselenide quantum emitters using hexagonal boron nitride wrinkles. *APL Photonics* **2020**, *5* (9).

(55) Yao, K.; Zhang, S.; Yanev, E.; McCreary, K.; Chuang, H. J.; Rosenberger, M. R.; Darlington, T.; Krayev, A.; Jonker, B. T.; Hone, J. C.; et al. Nanoscale Optical Imaging of 2D Semiconductor Stacking Orders by Exciton-Enhanced Second Harmonic Generation. *Advanced Optical Materials* **2022**, *10* (12).

(56) Shabani, S.; Darlington, T. P.; Gordon, C.; Wu, W.; Yanev, E.; Hone, J.; Zhu, X.; Dreyer, C. E.; Schuck, P. J.; Pasupathy, A. N. Ultralocalized Optoelectronic Properties of Nanobubbles in 2D Semiconductors. *Nano Lett* **2022**, *22* (18), 7401-7407.

(57) Jo, K.; Marino, E.; Lynch, J.; Jiang, Z.; Gogotsi, N.; Darlington, T. P.; Soroush, M.; Schuck, P. J.; Borys, N. J.; Murray, C. B.; et al. Direct nano-imaging of light-matter interactions in nanoscale excitonic emitters. *Nat Commun* **2023**, *14* (1), 2649.

(58) Paralikis, A.; Piccinini, C.; Madigawa, A. A.; Metuh, P.; Vannucci, L.; Gregersen, N.; Munkhbat, B. Tailoring polarization in WSe(2) quantum emitters through deterministic strain engineering. *NPJ 2D Mater Appl* **2024**, *8* (1), 59.

(59) Cai, H.; Rasmita, A.; He, R.; Zhang, Z.; Tan, Q.; Chen, D.; Wang, N.; Mu, Z.; Eng, J. J. H.; She, Y.; et al. Charge-depletion-enhanced WSe2 quantum emitters on gold nanogap arrays with near-unity quantum efficiency. *Nature Photonics* **2024**, *18* (8), 842-847.

(60) Kumar, S.; Brotóns-Gisbert, M.; Al-Khuzheyri, R.; Branny, A.; Ballesteros-Garcia, G.; Sánchez-Royo, J. F.; Gerardot, B. D. Resonant laser spectroscopy of localized excitons in monolayer WSe_2. *Optica* **2016**, *3* (8).

(61) von Helversen, M.; Greten, L.; Limame, I.; Shih, C.-W.; Schlaugat, P.; Antón-Solanas, C.; Schneider, C.; Rosa, B.; Knorr, A.; Reitzenstein, S. Temperature dependent temporal coherence of metallic-nanoparticle-induced single-photon emitters in a WSe2 monolayer. *2D Materials* **2023**, *10* (4).

(62) Piccinini, C.; Paralikis, A.; Neto, J. F.; Madigawa, A. A.; Wyborski, P.; Remesh, V.; Vannucci, L.; Gregersen, N.; Munkhbat, B. High-purity and stable single-photon emission in bilayer WSe(2) via phonon-assisted excitation. *Commun Phys* **2025**, *8* (1), 158.

(63) Paralikis, A.; Wyborski, P.; Metuh, P.; Gregersen, N.; Munkhbat, B. Tunable and Low-Noise WSe2 Quantum Emitters for Quantum Photonics. *PRX Quantum* **2025**, *6* (4).

(64) Novoselov, K. S.; Geim, A. K.; Morozov, S. V.; Jiang, D.; Zhang, Y.; Dubonos, S. V.; Grigorieva, I. V.; Firsov, A. A. Electric field effect in atomically thin carbon films. *Science* **2004**, *306* (5696), 666-669.

(65) Sun, B.; Grap, T.; Frahm, T.; Scholz, S.; Knoch, J. Role of electron and ion irradiation in a reliable lift-off process with electron beam evaporation and a bilayer PMMA resist system. *Journal of Vacuum Science & Technology B, Nanotechnology and Microelectronics: Materials, Processing, Measurement, and Phenomena* **2021**, *39* (5).

(66) Reina, A.; Jia, X.; Ho, J.; Nezich, D.; Son, H.; Bulovic, V.; Dresselhaus, M. S.; Kong, J. Large area, few-layer graphene films on arbitrary substrates by chemical vapor deposition. *Nano Lett* **2009**, *9* (1), 30-35.

(67) Lee, Y.; Bae, S.; Jang, H.; Jang, S.; Zhu, S. E.; Sim, S. H.; Song, Y. I.; Hong, B. H.; Ahn, J. H. Wafer-scale synthesis and transfer of graphene films. *Nano Lett* **2010**, *10* (2), 490-493.

(68) Castellanos-Gomez, A.; Buscema, M.; Molenaar, R.; Singh, V.; Janssen, L.; van der Zant, H. S. J.; Steele, G. A. Deterministic transfer of two-dimensional materials by all-dry viscoelastic stamping. *2D Materials* **2014**, *1* (1).

(69) Son, S.; Shin, Y. J.; Zhang, K.; Shin, J.; Lee, S.; Idzuchi, H.; Coak, M. J.; Kim, H.; Kim, J.; Kim, J. H.; et al. Strongly adhesive dry transfer technique for van der Waals heterostructure. *2D Materials* **2020**, *7* (4).

# Supporting information: Determination of the roles of strain and tearing in single photon emission from nanoindented $WSe_2$

Joseph Stage[1,*] , John Pierce Fix[1,2,†], Matthew C. Strasbourg[3], Amir Darabi[1], Torrey McLoughlin[1], Sheikh Parvez[1,2], Elijah Stuvland[1], Andrew R. Lingley[4], Thomas Darlington[5], and Nicholas J. Borys[1,6,*]

[1]Department of Physics, Montana State University, Bozeman, MT, 59717, USA

[2]Materials Science and Engineering, Montana State University, Bozeman, MT, 59717, USA

[3]Department of Mechanical Engineering, Columbia University, New York City, NY, 10027, USA

[4]Department of Electrical and Computer Engineering, Montana State University, Bozeman, MT, 59717, USA

[5]Molecular Foundry, Lawrence Berkeley National Laboratory, Berkeley, CA, 94720, USA

[6]Department of Physics and Astronomy, University of Utah, Salt Lake City, UT, 84109, USA

---

* J. S. and J. P. F. contributed equally to this work
* Corresponding author: nicholas.borys@utah.edu

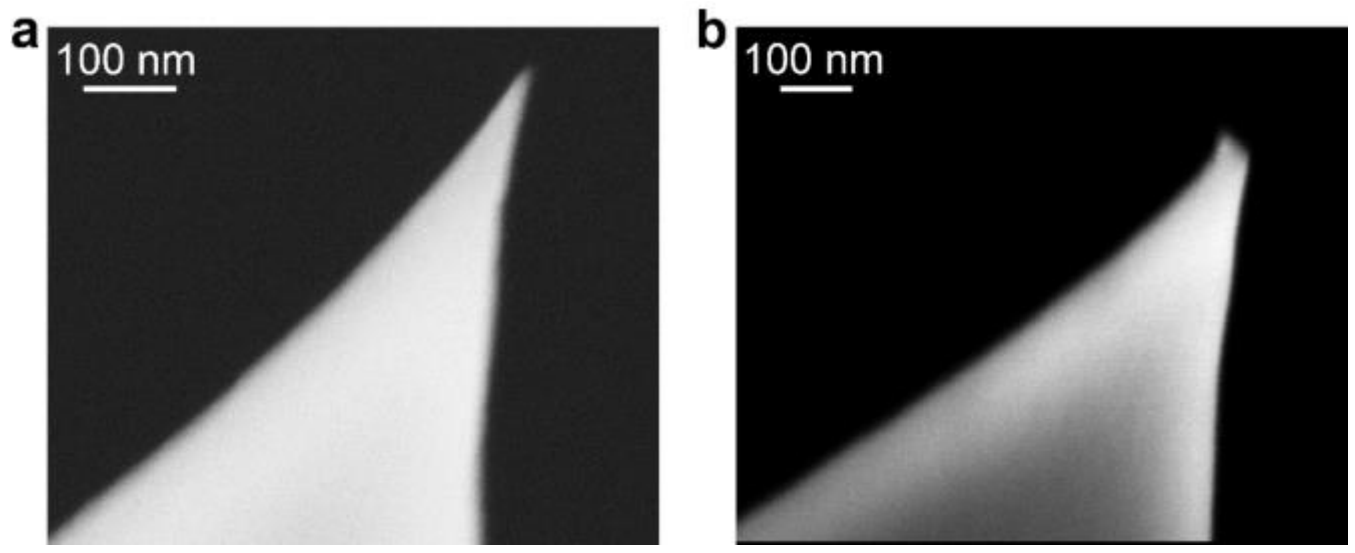

**Figure S1**: **Comparing new and dulled AFM tips via SEM. (a)** SEM images of a new Tap300Al-G AFM, k= 40 N/m probe before dulling. **(b)** SEM of the same AFM probe after dulling.

**Supporting information note 1:** Details indentation and indentation force calculation

To indent the 1L-$WSe_2$, we used Tap300Al-G AFM probes with a spring constant (k) of 40 N/m and a tip radius of 10 nm. During the nanoindentation process, the location of each indent and the movement of the sample stage into the probe (z-displacement) were set. This allows for control of the location and indentation force of each indent. At z-displacements over 1000 nm, a lateral displacement in the y-direction is introduced to keep the AFM tip in the same initial location throughout the indentation process. In the software, the y-displacement is set by inputting an angle. We can convert the z-displacement to indentation force by creating a calibration equation based on the indentation force (μN) that the software calculates as a function of z-displacement (see Figure S2). After 3050 nm of z-displacement, the reflected laser from the cantilever reaches the edge of the photodiode, and the indentation force calculation provided by the software becomes inaccurate. Motivating the need for a calibration equation to calculate indentation force from the z-displacement. Note that the software reports the z-displacement as negative values. The resulting equation from the linear fit is shown below.

$$\text{Indentation force } (\mu\text{N}) = -0.031 \left(\frac{\mu\text{N}}{\text{nm}}\right) * \text{z} + (-8.8\ \mu\text{N}) \tag{S1}$$

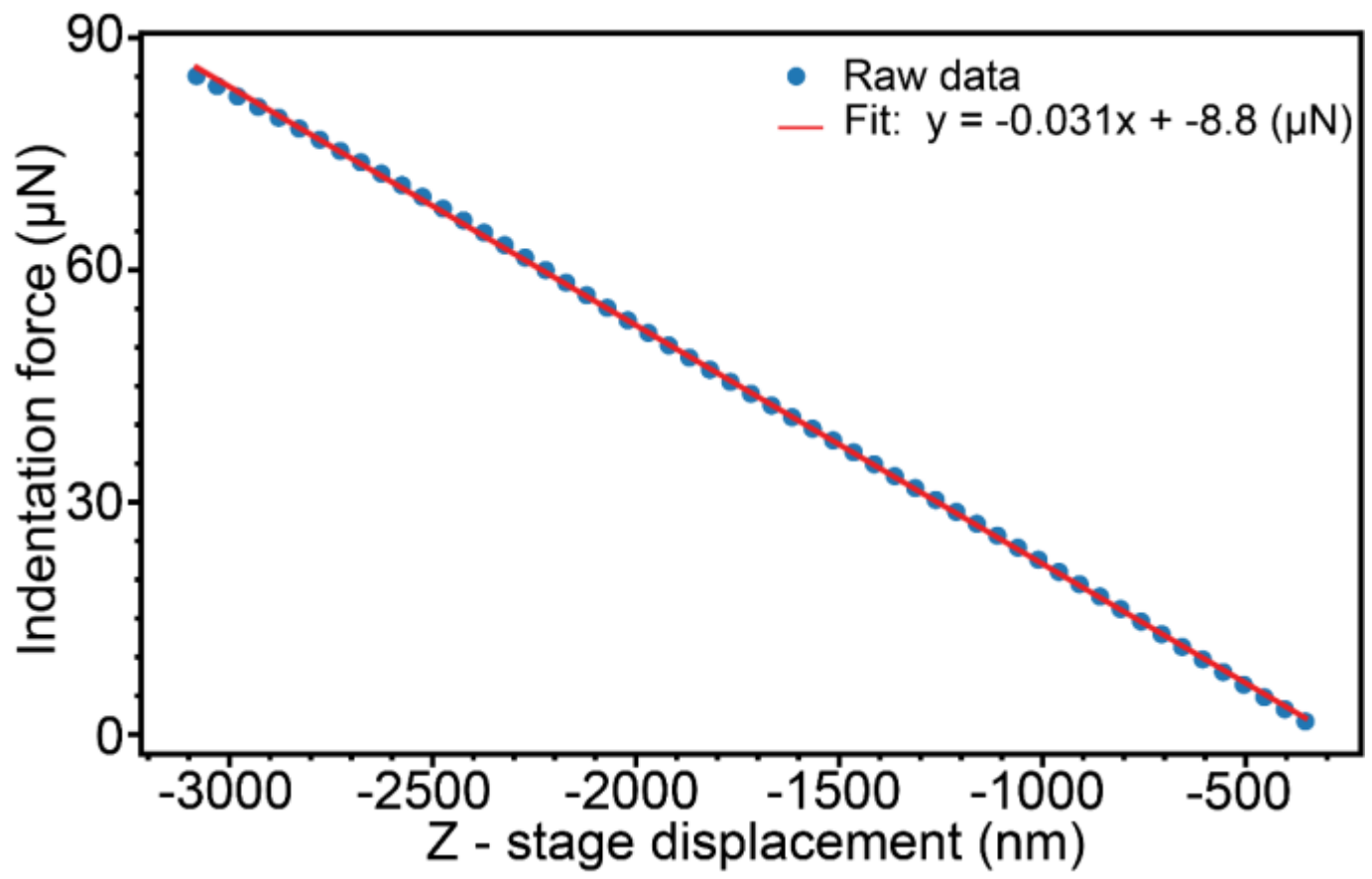

**Figure S2: Indentation force calibration.** Calculating an equation to convert z-displacement into indentation force (μN). The y-axis is the raw signal of the indentation force.

**Supporting information note 2:** Development of inversions with non-normal incidence Au evaporation

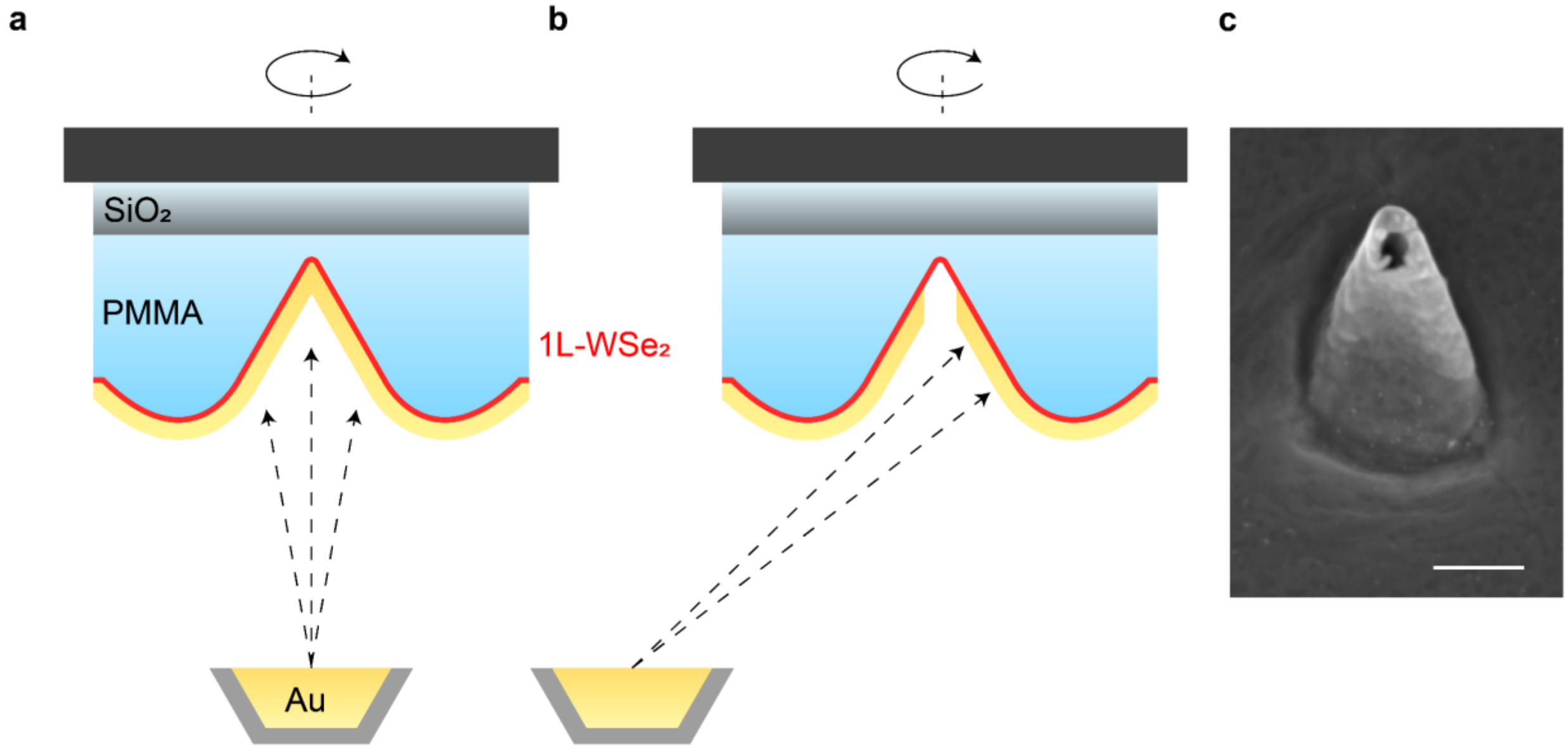

**Figure S3: Effect of non-normal incidence during inversion. (a)** Schematic showing perfect filling of an indent when a normal angle of incidence is used. **(b)** Schematic of a non-normal incidence evaporation leading to shadow masking at the apex of the indent/inversion. **(c)** SEM image of an inversion generated with non-normal incidence evaporation (scale bar = 200 nm).

During the inversion process, having the evaporated Au incident normal to the substrate ensures uniform filling of the indent as shown in Figure S3a. On the other hand, Figure S3b shows how non-normal evaporation can lead to shadow masking near the apex. SEM imaging of an inversion made with non-normal incidence evaporation can be seen in Figure S3c, where the apex is nominally void.

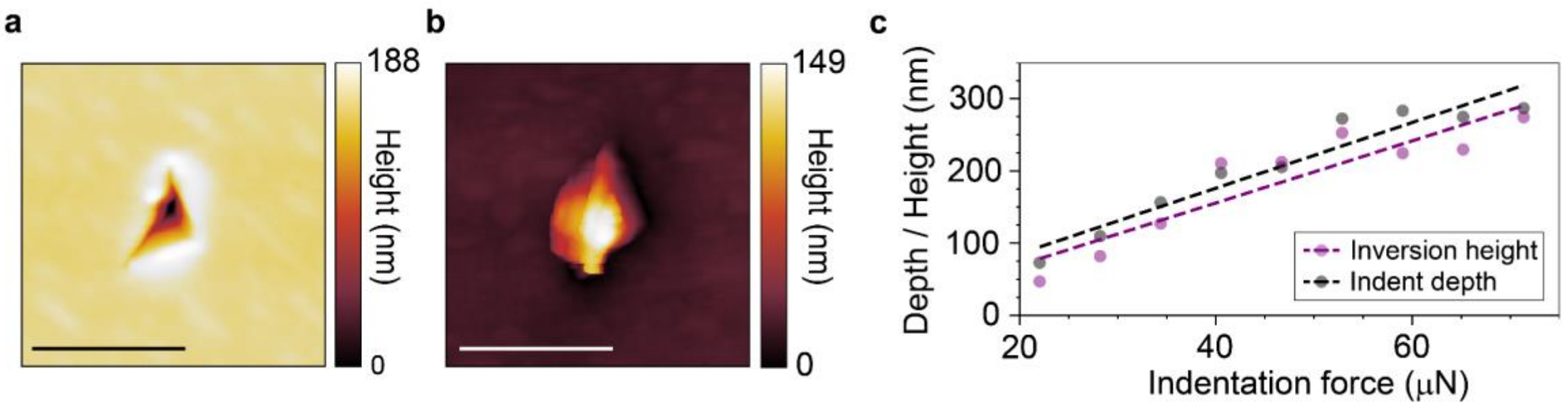

**Figure S4: Inversions are near replications of nanoindents. (a, b)** High-resolution AFM images of a 34 μN indent and inversion, respectively. **(c)** The depth of indents made using indentation forces ranging from 22 – 71 μN, and the heights of the corresponding inversions (scale bars = 500 nm).

**Supporting information note 3:** Sliding of the AFM tip from a misaligned contact angle

Due to the declination angle of the AFM cantilever with respect to the PMMA surface, a y-translation of the AFM stage is incorporated during indentation to maintain the location of the tip apex. The amount of y-translation required to maintain the tip position is prescribed in the Horiba AIST software through the “contact angle”, and the optimal value is a function of cantilever stiffness, declination angle, and substrate elastic modulus. Figure S5a schematically shows an example indentation with a contact angle of 0$^{o}$ (no y-translation of the AFM stage). As the AFM stage is brought up into the tip, sliding occurs due to cantilever deflection, leading to an enlarged side and an increased radius of curvature. We suspect another contributing factor to the enlarged apex radius of curvature of nanoinversions in comparison to the indenting AFM tip is the ~100x difference in elastic modulus between PMMA and 1L-$WSe_2$[1, 2]. Therefore, the strain applied at the apex results in much higher stress in the 1L-$WSe_2$ compared to PMMA, overcoming the adhesion at the interface, and softening the apex. Figure S5b represents an indent made with a perfectly aligned contact angle. Here, the stage moves laterally the same distance that the apex would be translated from the cantilever deflection. The indent generated is then a more accurate replication of the AFM tip geometry.

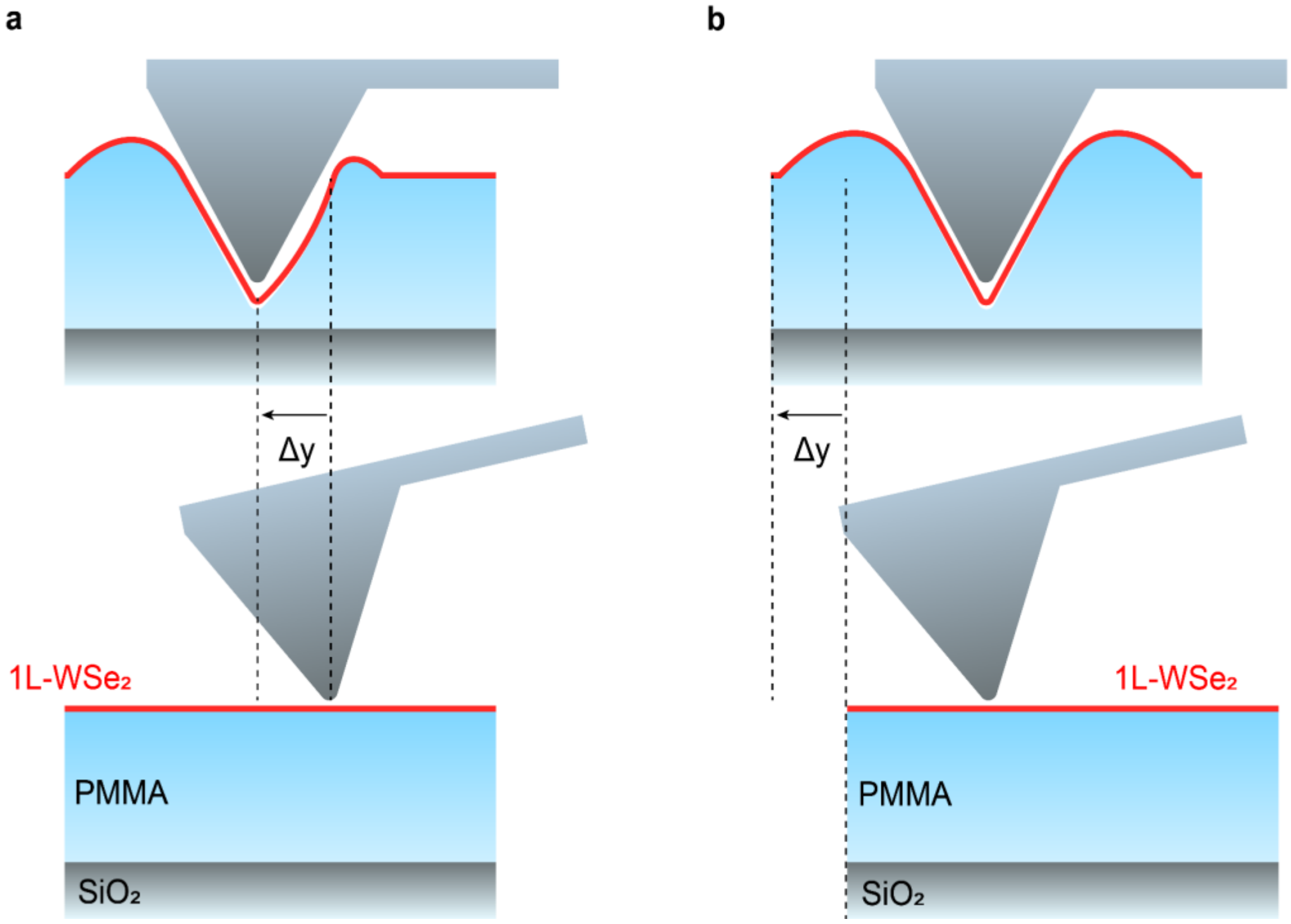

**Figure S5: Sliding of the AFM tip from a misaligned contact angle. (a)** Schematic representing an indent made with a contact angle of 0° (no y-translation of the AFM stage). **(b)** Schematic representing an indent made with a perfectly aligned contact angle.

**Supporting information note 4:** Yaw tilt of the AFM cantilever generating an asymmetric buildup region

Yaw tilt is described as a rotation of the cantilever with respect to the PMMA surface when looking at the cantilever head-on. This can occur from small misalignments in the mounting of the AFM tip (dust under the Si, AFM tip fabrication issues) or from the substrate surface being tilted. During indentation, yaw tilt leads to an asymmetry in the build region. The side that the tip is rotated into will generate a larger buildup region since PMMA is displaced in that direction (seen in Figure S6).

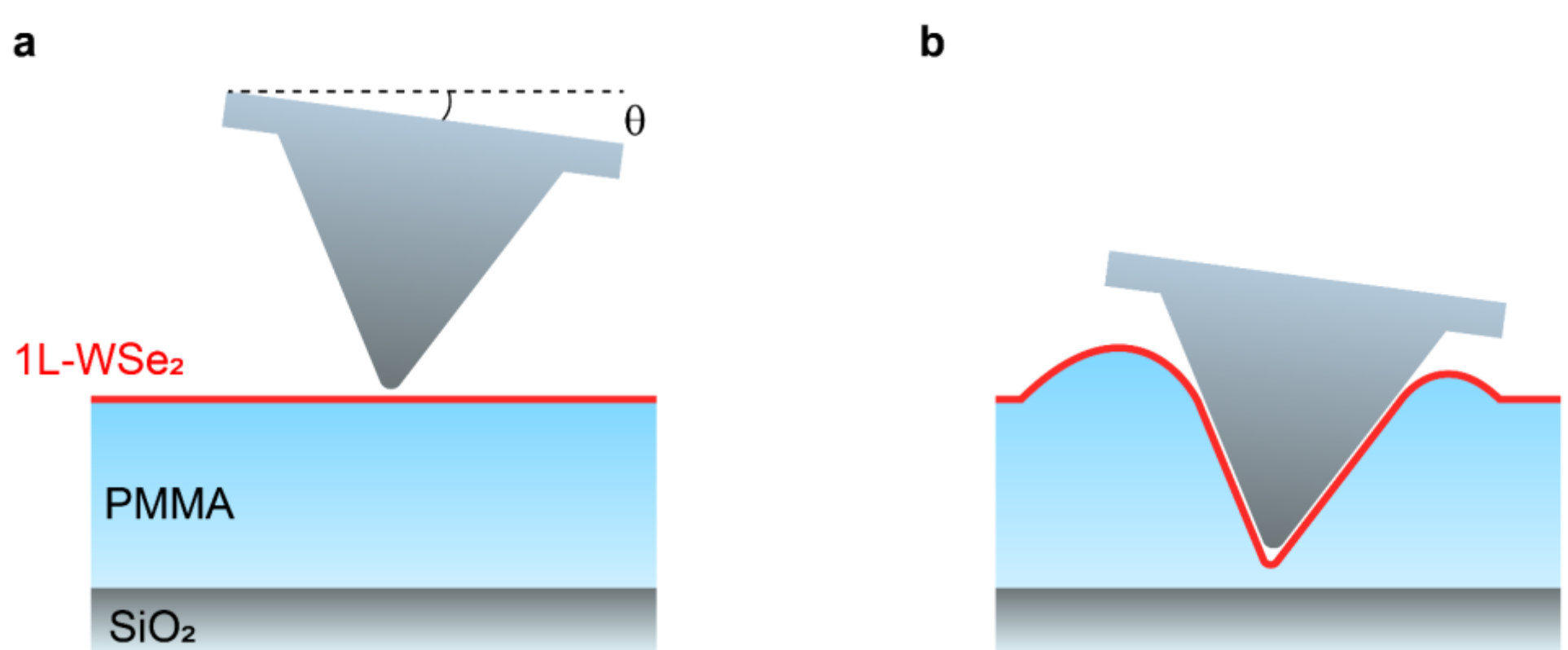

**Figure S6: Yaw tilt generating an asymmetric buildup region. (a)** Schematic of an AFM tip with yaw tilt (θ) positioned above the PMMA surface before indentation. **(b)** Schematic of indentation where the AFM cantilever has a yaw tilt leading to an asymmetric buildup region.

**Supporting information note 5:** Tearing in indented materials from dulled AFM tips with different y-displacements (angles)

To see the effect of the y-displacement on the indents, defined by the contact angle setting in the AFM software, we made two inverted indent arrays with a constant indentation force of 53 μN, shown in Figure S7. The array shown in Figure S7a was made with a dulled Tap300Al-G AFM, k = 40 N/m probe. As the y-displacement increases and limits the effect of the AFM tip slipping forward during the indentation process, the base of the inverted indent gets smaller, which means the AFM tip stays in the initial location of the tip (see Figure S7a). This is the behavior seen from 0° to 24°. In contrast, from 26° to 60° the y-displacement causes the natural tip slipping to increase, resulting in the inverted indent base to elongate and appear stretched. High-resolution SEM images are taken of the inverted indents, and two examples are shown in Figures S7b-c. In Figures S7b-c, tears are observed in the 1L-$WSe_2$. Tears like the ones seen in Figure S7b-c are observed in all the other inverted indents in Figure S7a. In contrast, the inverted indents in Figures S7d-e are made with the same probe before dulling, but they do not show tearing.

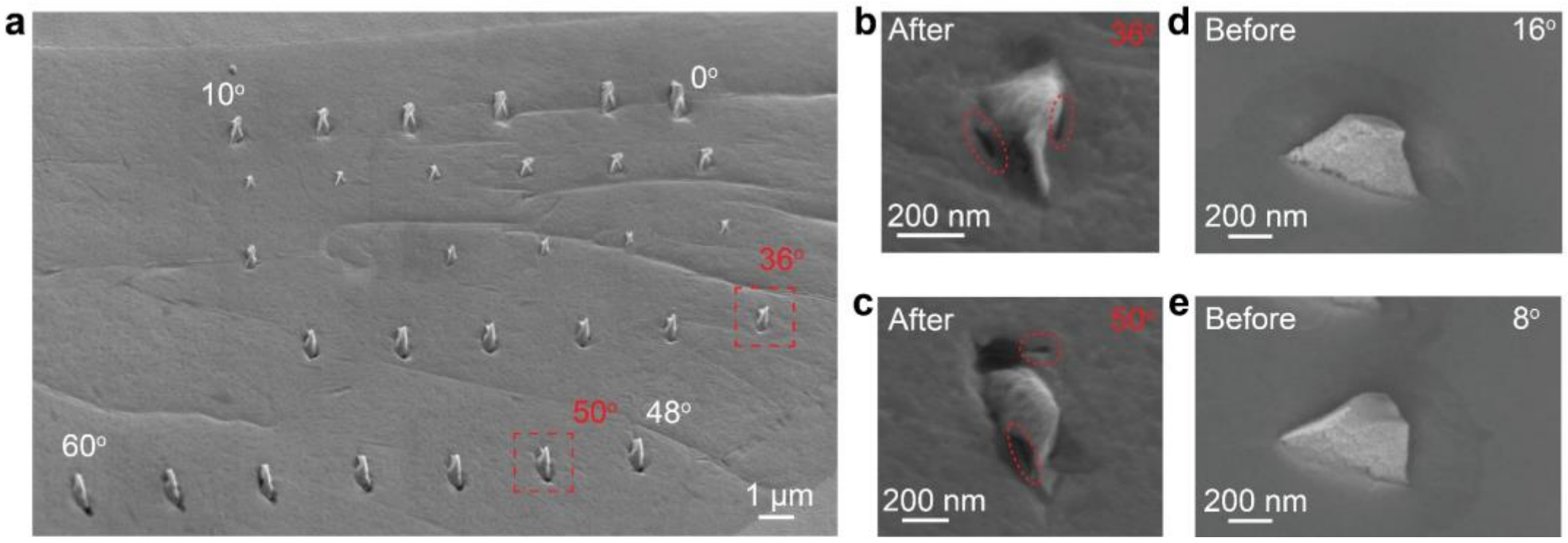

**Figure S7: Two inverted 1L-$WSe_2$ indent arrays made with the same AFM probe before and after dulling**. The same indentation force was used (53 μN), and the angle was changed from 0 to 60° in steps of 2° for both arrays of indents. **(a)** SEM image of an array of inverted indents made with a dulled Tap300Al-G, k = 40 N/m AFM probe. The angles set the y-displacement of the sample stage. **(b, c)** SEM of two indents outlined by red squares in (a) made with angles of 36° and 50° respectively. The red ovals mark tears in the 1L-$WSe_2$. **(d, e)** SEM of inverted indents made with a fresh AFM probe, meaning the probe was not dulled. The indents were made with 16° (d) and 8° (e) and do not show signs of tearing as seen in (b) and (c).

**Supporting information note 6:** Limitations of thin plate theory for modeling the strain in indents/inversions

To demonstrate the limitations of using a simple, thin plate model to analyze the strain in the indentation, we report the strain in a thin plate (1L-$WSe_2$) conformed to the geometry of a ~53 μN nanoinversion using the Föppl–von Kármán equations[3] in Figure S8. The results of this model represent a hypothetical system with excessively large strain values, far beyond the mechanical failure point of a 1L-$WSe_2$ crystallite[2, 4]. Two important considerations of this model, which limit its applicability to analyzing this system, are that the 1L-$WSe_2$ is assumed to be fixed and is not allowed to slip, and that the PMMA surface is assumed to be rigid and not allowed to shear. As a result of these assumptions, the lateral motion of the 1L-$WSe_2$ surrounding the nano-indentation/inversion site is neglected, so strains in excess of 200% are predicted. This model strongly suggests that during the nanoindentation process, the flat 1L-$WSe_2$ surrounding the nanoindent experiences a strain gradient from either lateral motion of the 1L-$WSe_2$ along the PMMA surface or shearing of the PMMA underneath the 1L-$WSe_2$, limiting the amount of strain that can be ultimately generated by the indent. Further, these results could provide potential explanations for the tearing of nanoindents. For example, if the slipping or shearing of the 1L-$WSe_2$ and PMMA is inhibited by environmental or physical constraints, tearing may be likely to occur. Overall, more accurate modeling of the strain likely requires consideration of the dynamics of the indentation process itself, including accounting for things such as slippage of the 2D material as well as the viscoelastic/viscoplastic deformation of the PMMA.

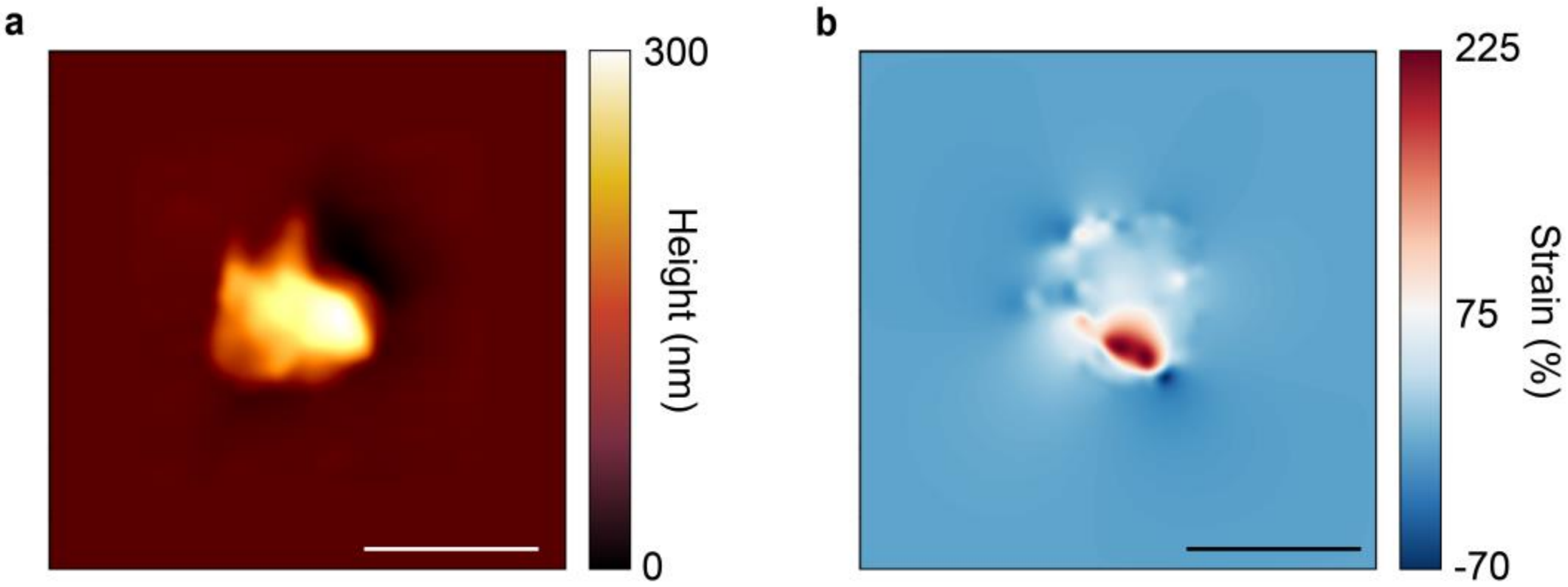

**Figure S8: Föppl–von Kármán strain modeling. (a)** Smoothed AFM image of the ~53 μN nanoinversion in Batch 2. **(b)** Hypothetical strain modelling results from solving the Föppl–von Kármán equations for a thin plate given the geometry seen in (a). Scale bars for (a) and (b) are 500 nm.

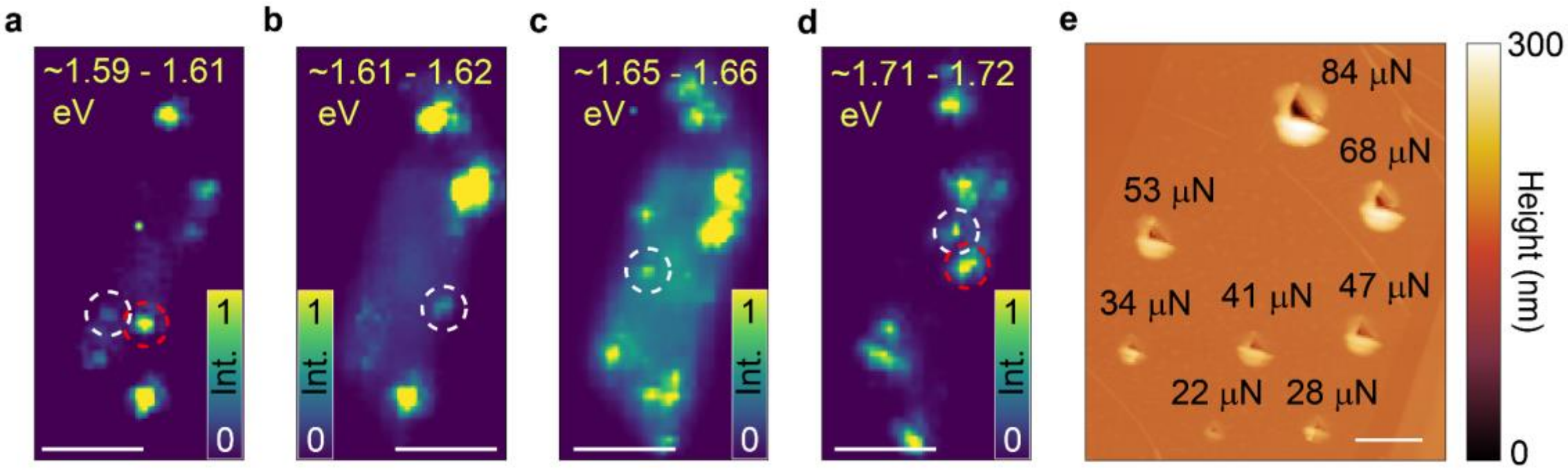

**Figure S9: Locating indentation sites in Batch 1. (a)** Hyperspectral image integrated from ~1.59 – 1.61 eV used to identify the 34 and 41 μN indents in white and red, respectively. **(b)** Hyperspectral image integrated from ~1.61 – 1.62 eV used to identify the 47 μN indent. **(c)** Hyperspectral image integrated from ~1.65 – 1.66 eV used to identify the 53 μN indent. **(d)** Hyperspectral image integrated from ~1.71 – 1.72 eV used to identify the 68 and 84 μN indents in red and white, respectively (scale bars = 5 μm). **(e)** AFM image of the indents in Batch 1, ranging in indentation force from 22 – 84 μN (scale bar = 1 μm).

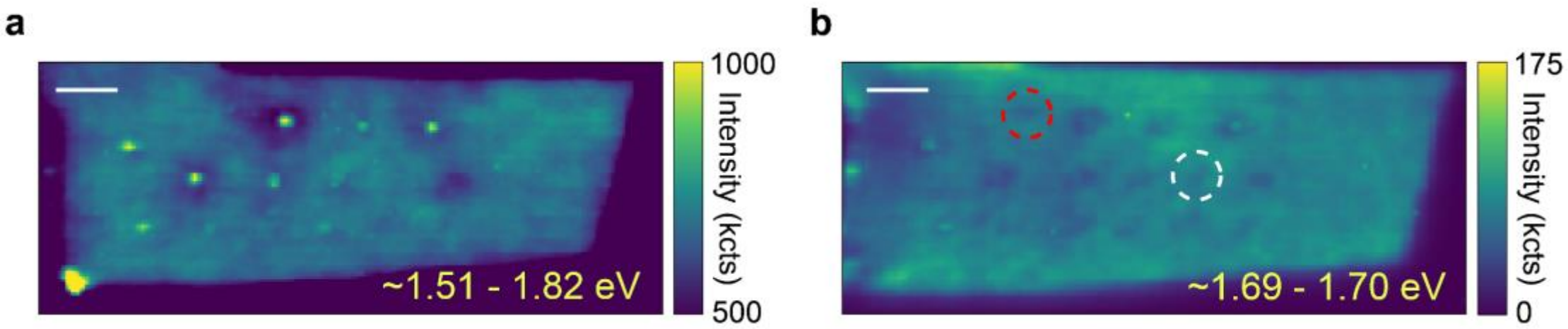

**Figure S10: Locating indentation sites in Batch 2. (a)** Hyperspectral image integrated from 1.51 – 1.82 eV. **(b)** Hyperspectral image integrated from 1.69 – 1.70 eV used to identify the 28 and 65 μN indents in white and red, respectively (scale bar = 10 μm).

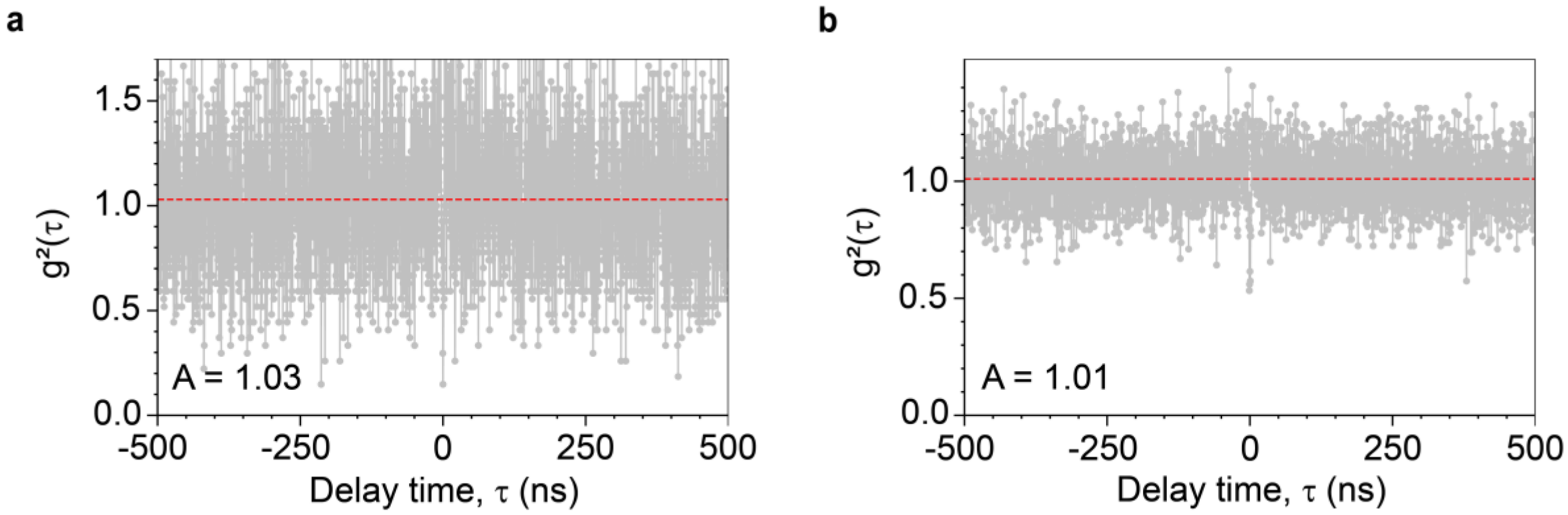

**Figure S11: Extended delay ranges from manuscript Figure 4. (a)** Extended delay second-order correlation function of manuscript Figure 4c (Batch 1). **(b)** Extended delay second-order correlation function of manuscript Figure 4d (Batch 2). The red dashed lines represent the background level from the second-order correlation function fit used to extract the $g^2(0)$ values reported in the main manuscript. The larger noise seen in (a) is indicative of the reduced SPE intensity from the Batch 1 (torn) nanoindents.

**Supporting information note 7:** Properties of the emitter ensemble at each site in Batch 3

4 K photoluminescence (PL) taken at each indentation site of Batch 3 was used to investigate if properties of the emitter ensemble had a dependence on indentation force. Similar to previous reports, in Figures S12a-d, we note no trend of integrated PL intensity, average emitter energy, highest emitter energy, or lowest emitter energy to indentation force (strain)[5, 6]. These metrics were extracted using an average PL spectra from a 120 s PL time series, both before and after, graphite was added to quench excessive background emission[7]. Determination of average, highest, and lowest emitter energy was done using the emitters identified from the counting protocol.

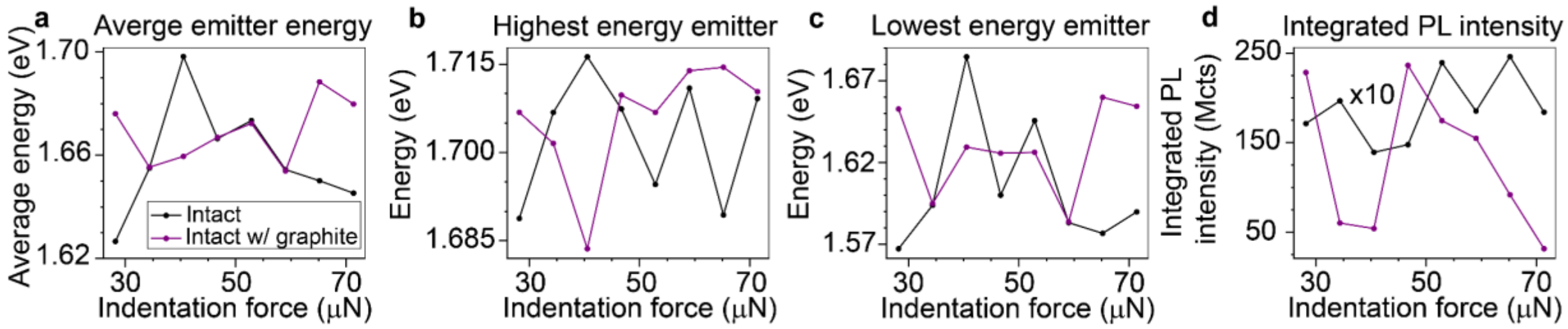

**Figure S12: Emitter ensemble properties of each site in Batch 3. (a)** Average energy of the emitter ensemble at each indentation site. **(b, c)** The lowest and highest energy emitter of the ensemble at each site, respectively. **(d)** Integrated PL intensity of all emission (1.575 – 1.725 eV) at each indentation site.

**Supporting information note 8:** AFM characterization of the graphite overlaid on the indented 1L-$WSe_2$

Figure S13 shows the topography of the graphite layer on top of the indented 1L-$WSe_2$ discussed in the main text. Significant bulging of the graphite at each indentation site occurs, indicating that the graphite is suspended over the indent and supported by the surrounding buildup regions. The heights of the bulges increase with indentation force in the same manner as the heights of the buildup regions created by the indentation process.

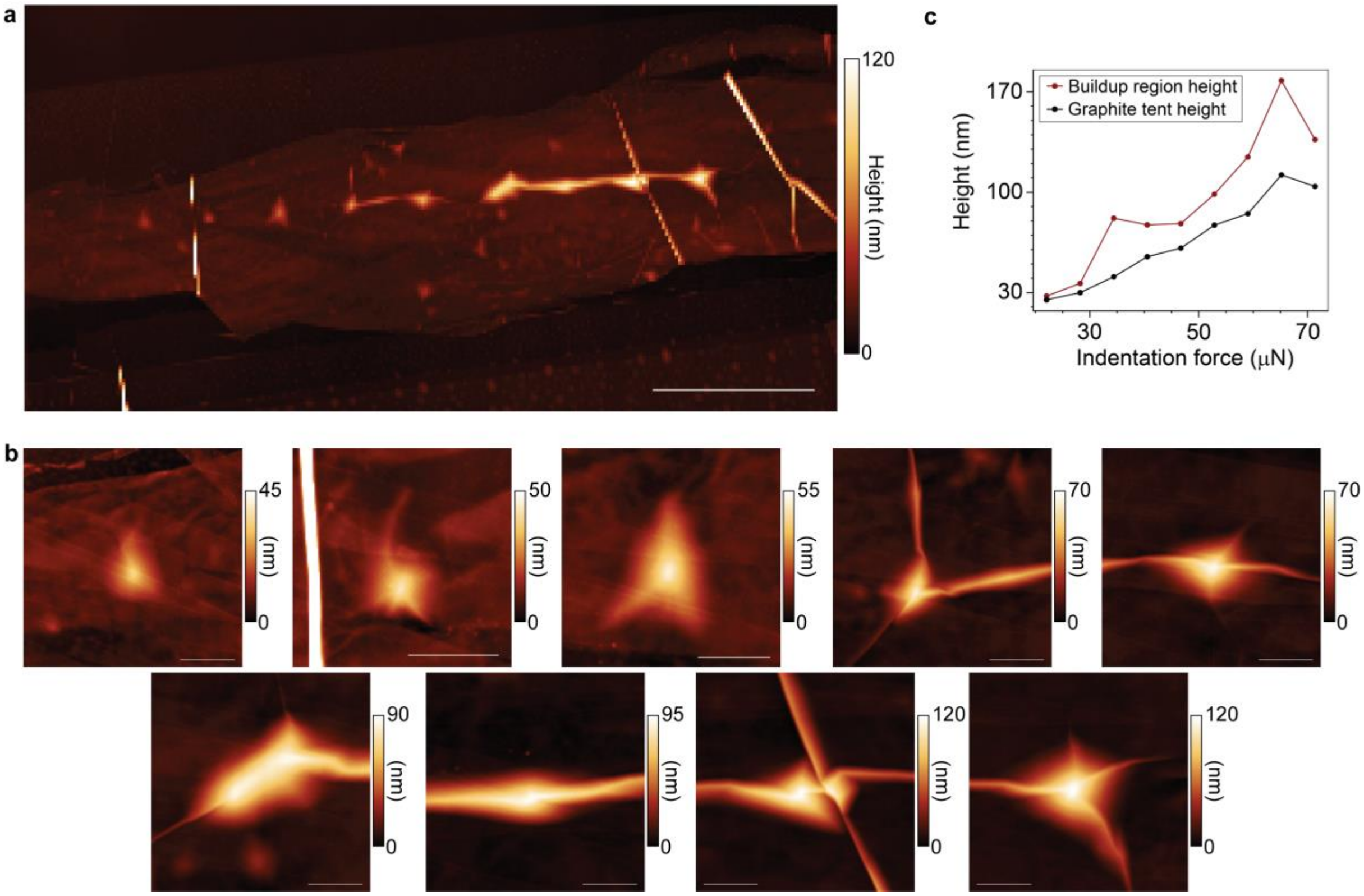

**Figure S13: AFM characterization of graphite overlaid on the indented 1L-$WSe_2$. (a)** Large area scan showing that the graphite fully coveres the indented 1L-$WSe_2$.Scale bar: 10 μm. **(b)** High-resolution AFM scans of each indent. Scale bars: 1 μm. (c) Heights of the graphite bulges compared to the heights of the buildup regions.

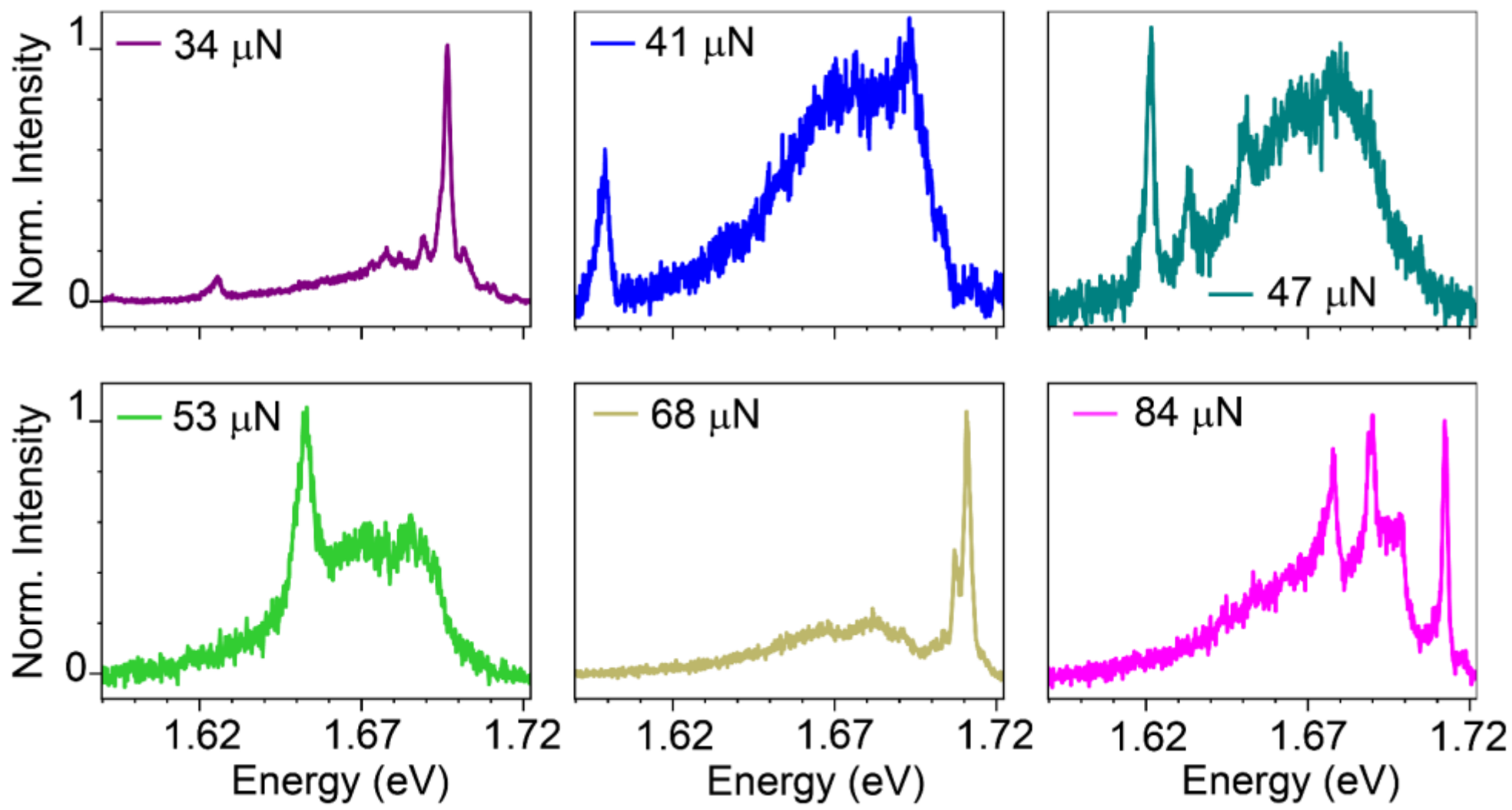

**Figure S14: PL spectra of each site in Batch 1.** PL spectra of indentation sites made with indentation forces ranging from 34 – 84 μN**.**

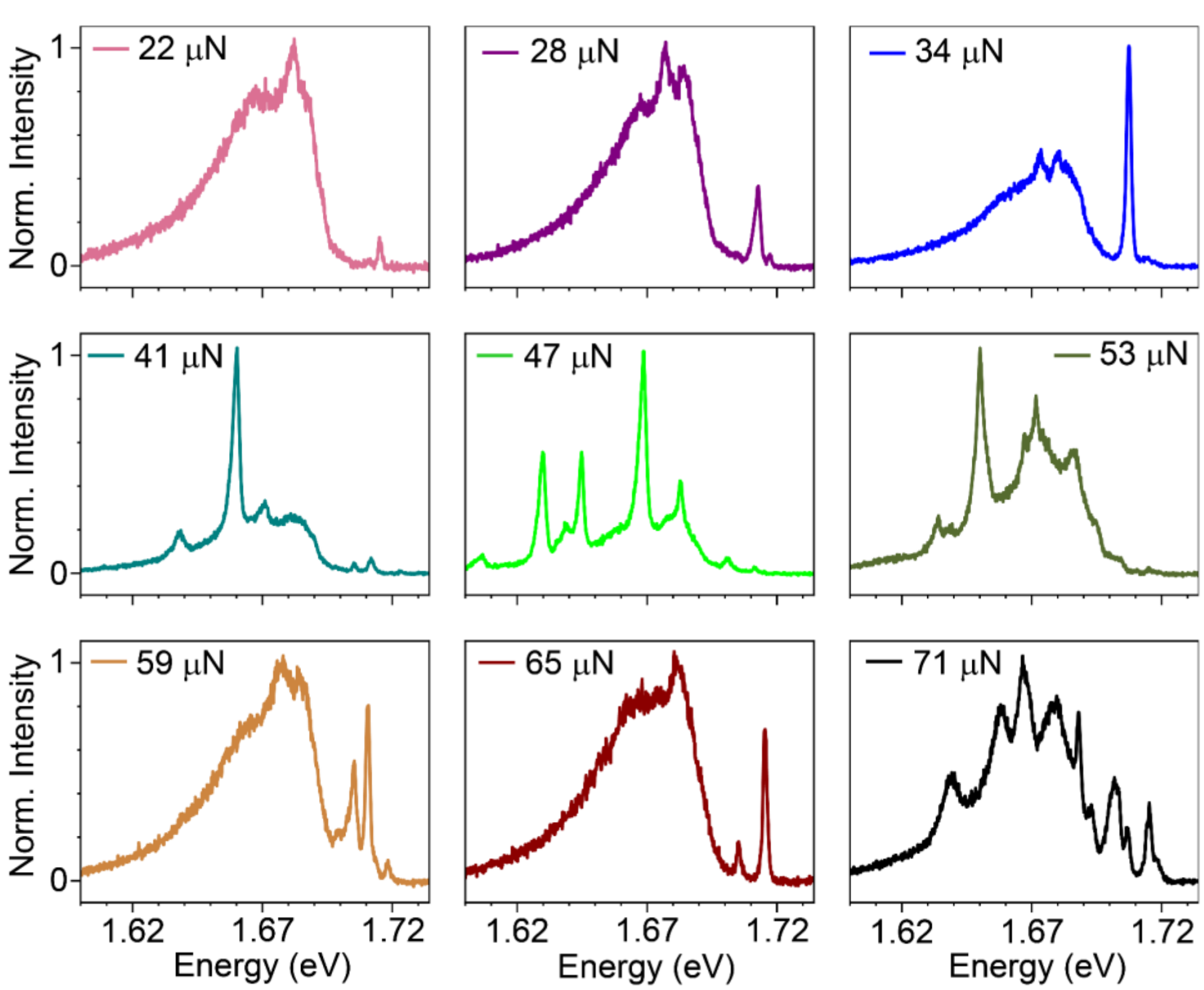

**Figure S15: PL spectra of each site in Batch 2.** PL spectra of indentation sites made with indentation forces ranging from 22 – 71 μN.

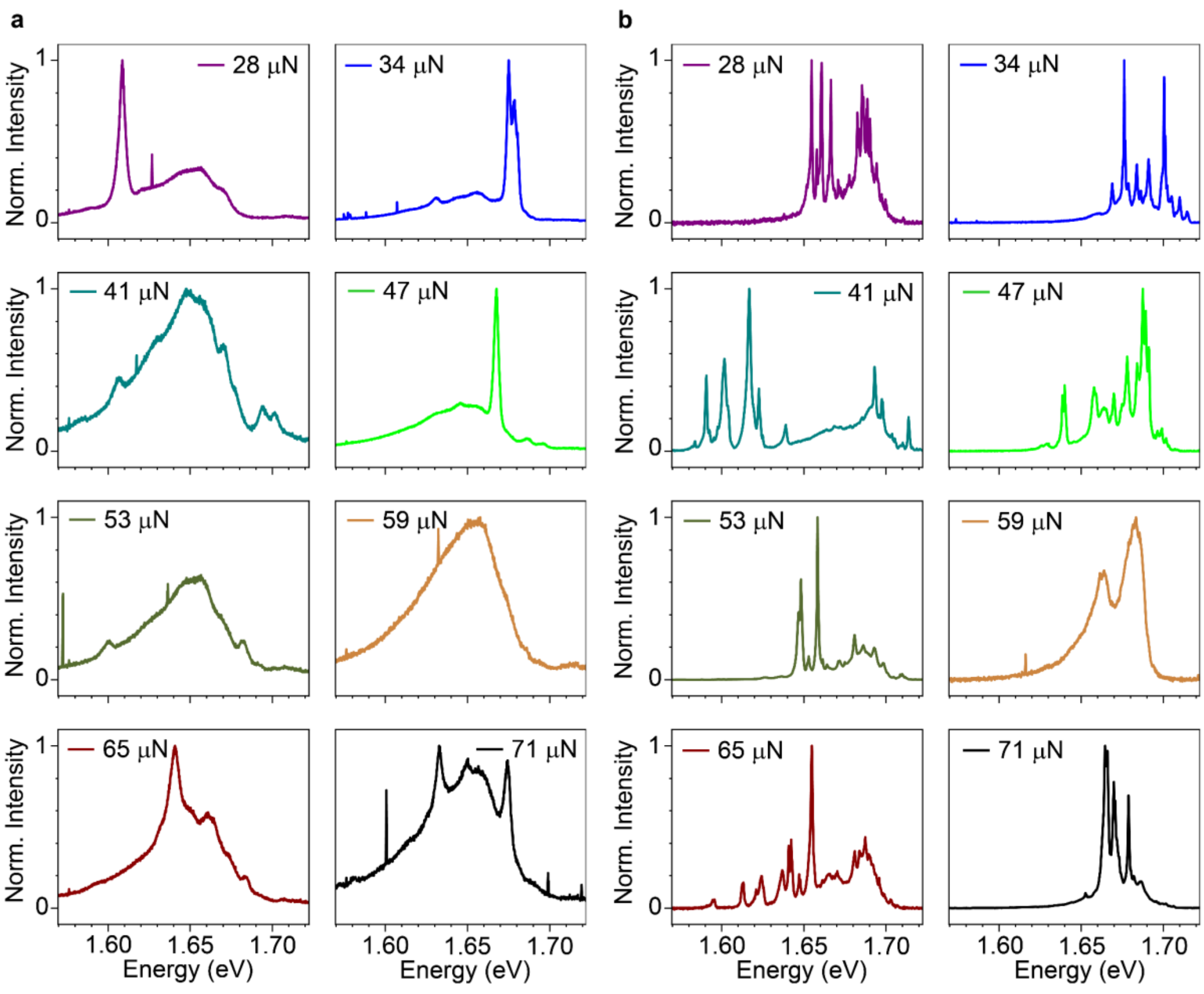

**Figure S16: Average PL spectra of each site in Batch 3.** Average PL spectra of indentation sites made with indentation forces ranging from 28 – 71 μN. **(a)** Average PL spectra from a 120 s PL time-series without graphite. **(b)** Average PL spectra from a 120 s PL time-series with graphite.

**Supporting information note 8:** Total area and buildup area of each indentation site

To understand how the total area and buildup area of each indentation site increase with added indentation force, we have taken high-resolution AFM images and calculated the total and buildup areas as seen in Figure S17. Both the total and buildup areas increase linearly with added indentation force in all samples.

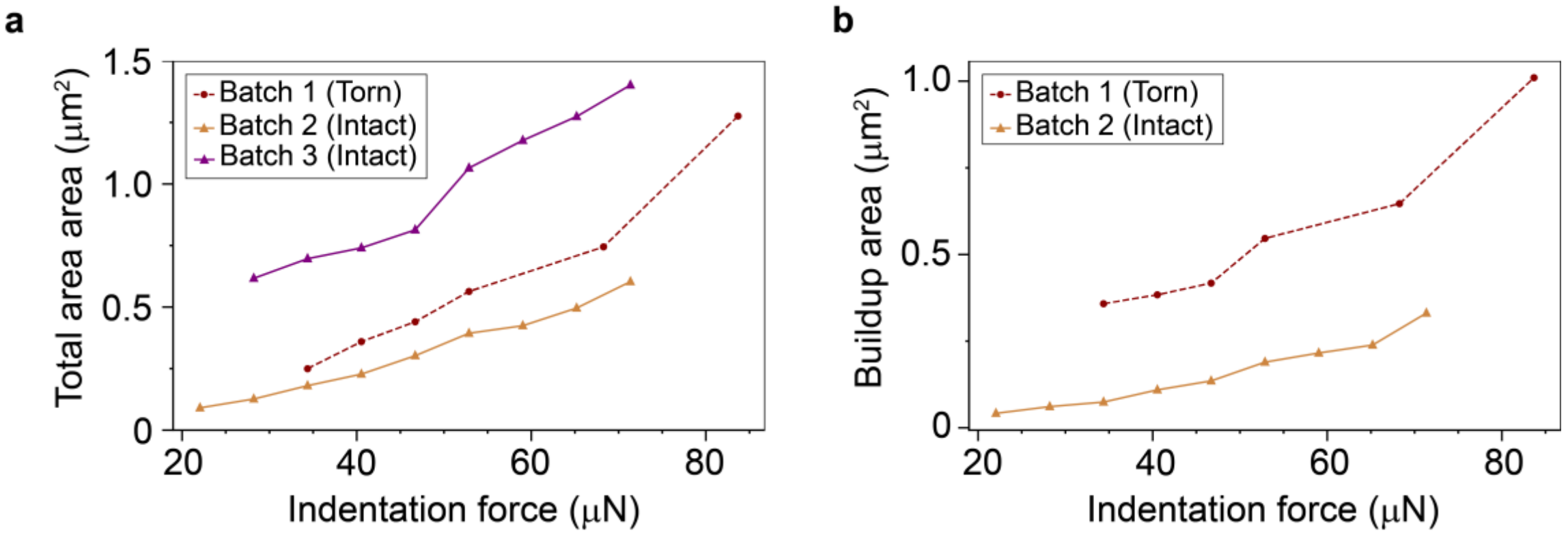

**Figure S17: AFM obtained areas of each indentation site. (a)** Total area of each indentation site, including the buildup region. **(b)** Area of the buildup region for Batch 1 and Batch 2. Areas are obtained using high-resolution AFM imaging.

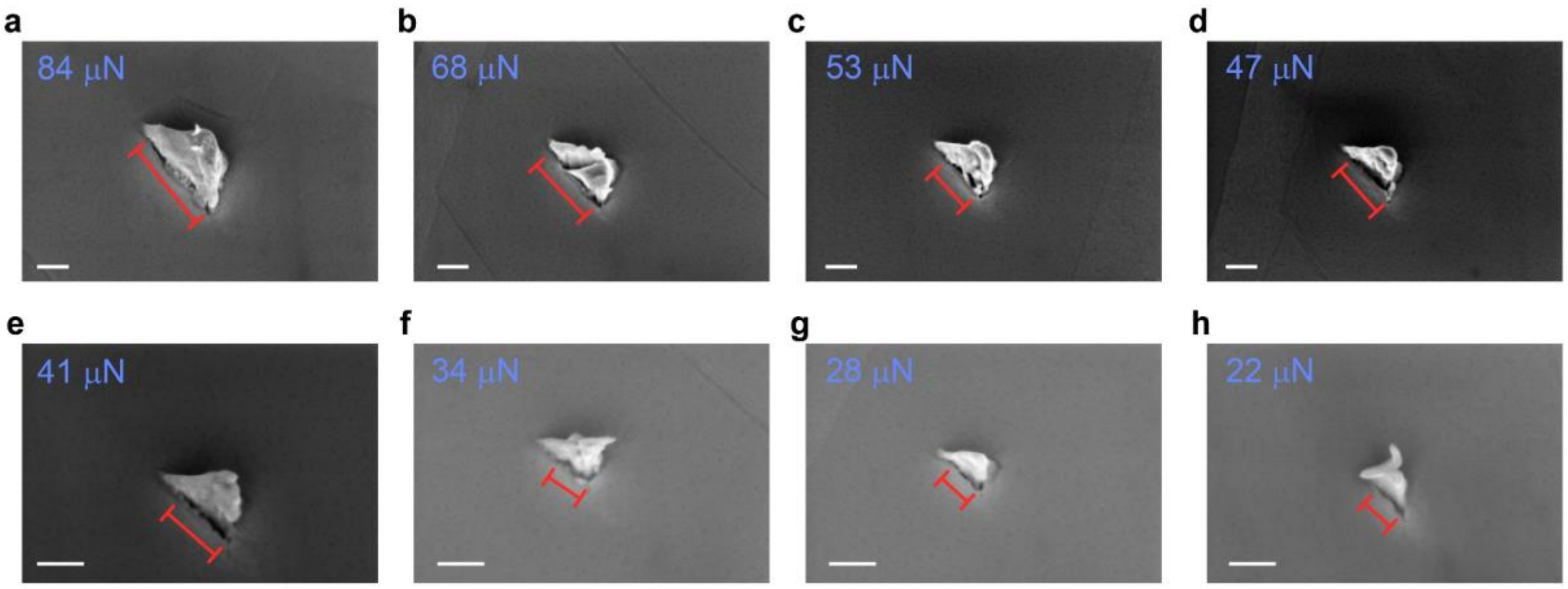

**Figure S18: Tear length of Batch 1 indents using SEM imaging. (a-h)** SEM images of Batch 1 inversions. Torn regions are highlighted using a red line, and the indentation force used to generate each indent is listed in blue (scale bars = 200 nm).

**References**

(1) Stafford, C. M.; Harrison, C.; Beers, K. L.; Karim, A.; Amis, E. J.; VanLandingham, M. R.; Kim, H. C.; Volksen, W.; Miller, R. D.; Simonyi, E. E. A buckling-based metrology for measuring the elastic moduli of polymeric thin films. *Nat Mater* **2004**, *3* (8), 545-550.

(2) Falin, A.; Holwill, M.; Lv, H.; Gan, W.; Cheng, J.; Zhang, R.; Qian, D.; Barnett, M. R.; Santos, E. J. G.; Novoselov, K. S.; et al. Mechanical Properties of Atomically Thin Tungsten Dichalcogenides: $WS_2$, $WSe_2$, and $WTe_2$. *ACS Nano* **2021**, *15* (2), 2600-2610.

(3) Darlington, T. P.; Krayev, A.; Venkatesh, V.; Saxena, R.; Kysar, J. W.; Borys, N. J.; Jariwala, D.; Schuck, P. J. Facile and quantitative estimation of strain in nanobubbles with arbitrary symmetry in 2D semiconductors verified using hyperspectral nano-optical imaging. *J Chem Phys* **2020**, *153* (2), 024702.

(4) Ben-Smith, A.; Choi, S. H.; Boandoh, S.; Lee, B. H.; Vu, D. A.; Nguyen, H. T. T.; Adofo, L. A.; Jin, J. W.; Kim, S. M.; Lee, Y. H.; et al. Photo-oxidative Crack Propagation in Transition Metal Dichalcogenides. *ACS Nano* **2024**, *18* (4), 3125-3133.

(5) Rosenberger, M. R.; Dass, C. K.; Chuang, H. J.; Sivaram, S. V.; McCreary, K. M.; Hendrickson, J. R.; Jonker, B. T. Quantum Calligraphy: Writing Single-Photon Emitters in a Two-Dimensional Materials Platform. *ACS Nano* **2019**, *13* (1), 904-912.

(6) Yücel, O.; Yagodkin, D.; Kirchhof, J. N.; Yu, Y.; Kumar, A. M.; Dewambrechies, A.; Kovalchuk, S.; Bolotin, K. I. Strain activation of localized states in $WSe_2$. *2D Materials* **2025**, *12* (3).

(7) Stevens, C. E.; Chuang, H. J.; Rosenberger, M. R.; McCreary, K. M.; Dass, C. K.; Jonker, B. T.; Hendrickson, J. R. Enhancing the Purity of Deterministically Placed Quantum Emitters in Monolayer $WSe_2$. *ACS Nano* **2022**, *16* (12), 20956-20963.